\documentclass[aps,twocolumn,pra,showpacs,amsmath,amssymb,showkeys,10pt]{revtex4-2}
\usepackage{graphicx}
\usepackage{epstopdf}
\pdfoutput=1
\usepackage{braket}
\usepackage{hyperref}
\usepackage{longtable}
\newcommand{\tabincell}[2]{\begin{tabular}{@{}#1@{}}#2\end{tabular}}

\begin{document}

	\title{Using a modified version of the Tavis--Cummings--Hubbard model to simulate the formation of neutral hydrogen molecule}

	\author{Miao Hui-hui}
	\affiliation{Faculty of Computational Mathematics and Cybernetics, Lomonosov Moscow State University, Vorobyovy Gory 1, Moscow, 119991, Russia}

	\author{Ozhigov Yuri Igorevich}
	\email[Email address: ]{ozhigov@cs.msu.ru}
	\affiliation{Faculty of Computational Mathematics and Cybernetics, Lomonosov Moscow State University, Vorobyovy Gory 1, Moscow, 119991, Russia\\K. A. Valiev Institute of physics and technology, Russian Academy of Sciences, Nakhimovsky Prospekt 32-a, Moscow, 117218, Russia}

	\date{\today}

	\begin{abstract}	
	A finite-dimensional chemistry model with two two-level artificial atoms on quantum dots positioned in optical cavities, called the association--dissociation model of neutral hydrogen molecule, is described. The initial circumstances that led to the formation of the synthetic neutral hydrogen molecule are explained. In quantum form, nuclei's mobility is portrayed. The association of atoms in the molecule is simulated through a quantum master equation, incorporating hybridization of atomic orbitals into molecular --- depending on the position of the nuclei. Consideration is also given to electron spin transitions. Investigated are the effects of temperature variation of various photonic modes on quantum evolution and neutral hydrogen molecule formation. Finally, a more precise model including covalent bond and simple harmonic oscillator (phonon) is proposed.
	\end{abstract}


	\maketitle

	\section{Introduction}
	\label{sec:Introduction}

	The ability of supercomputers to simulate restricted molecular structures within the framework of "quantum chemistry" --- stationary states of molecules, has increased interest in mathematical modelling of natural phenomena, particularly predictive modelling of chemistry. This interest has recently been sparked by various theoretical papers, including those by \cite{Wang2021, McClean2021, Claudino2022}. Even before the construction of the full-scale quantum computer \cite{McClean2021}, quantum approaches open new perspectives for effectively modelling well-known effects and witnessing fundamentally novel phenomena in chemistry, compared to classical approaches. The modelling of hydrogen-related chemical reactions, particularly the formation and decomposition (reactions of association and dissociation, respectively) of the cation H$_2^+$ and neutral hydrogen molecule H$_2$, is one of the main objectives of chemical modelling. The construction of large molecular structures, notably biomacromolecules like proteins and deoxyribonucleic acid, necessitates an understanding of hydrogen chemical processes. In works \cite{Zhu2020, Afanasyev2021}, a thorough simulation of association and dissociation of the cation H$_2^+$ is put forward. The association reaction of the neutral hydrogen molecule H$_2$ in open Markovian systems is the focus of this paper.
	
	The quantum electrodynamics (QED) model, which presents a distinct physical paradigm for examining interaction between light and matter, is a fundamental contribution to this paper. In this paradigm, fields (of cavities) are related to impurity two- or multi-level systems, which are typically referred to as atoms. We must utilize models resembling finite dimensional cavity QED models for "dynamical chemistry" because the description of the field is the major area of difficulty. The ultrastrong-coupling \cite{Haroche2013, Gu2017, Kockum2019, Frisk2019, Forn-Diaz2019} (USC) of light and matter(e.g., a cavity mode and a natural or artificial atom, respectively) occurs when their coupling strength $g$ becomes comparable to the atomic ($\omega_a$) or cavity ($\omega_c$) frequencies. More specifically, the USC regime occurs when $\eta=max\left(\frac{g}{\hbar\omega_c},\frac{g}{\hbar\omega_a}\right)$ is within the range $\left[0.1, 1\right)$. The quantum Rabi model (QRM) \cite{Rabi1936, Rabi1937} is the fundamental model for USC of a single two-level atom in a single-mode cavity. The Dicke \cite{Dicke1954} and Hopfield \cite{Hopfield1958} models are two examples of its multi-atom or multi-mode generalizations. Deep strong coupling (DSC) is a common term used to describe the regime $\eta\geq 1$ \cite{Casanova2010}. The more straightforward strong coupling (SC) model --- Jaynes--Cummings model (JCM) \cite{Jaynes1963} can be used to replace these models for USC when $\eta<0.1$. The JCM depicts the dynamics of a two-level atom in an optical cavity, interacting with a single-mode field inside it. Its generalization --- the Tavis--Cummings model (TCM) \cite{Tavis1968} depicts the dynamics of a collection of $N$ two-level atoms in an optical cavity. The Jaynes--Cummings--Hubbard model (JCHM) and Tavis--Cummings--Hubbard model (TCHM) \cite{Angelakis2007} are generalizations of the JCM and TCM to multiple cavities coupled by an optical fibre. Due to the fact that SC is often simpler to realize in an experiment than USC and DSC, we modified these SC models in this paper to fulfil the needs of chemical reaction simulation. As finite-dimensional QED models, these models and their modifications are valuable because they enable us to describe a very complex interaction between light and matter. Among these models, the optical cavity --- Fabry--Pérot resonator is the most significant form, where atoms are held in place by optical tweezers. Many studies have been conducted recently in the field of JCM and its modifications, including those on phase transitions \cite{Wei2021, Prasad2018}, the search for metamaterials \cite{Guo2019}, quantum many-body phenomena \cite{Smith2021}, the realization of Grover search algorithm \cite{Kulagin2022}, quantum gates \cite{OzhigovYI2020, Dull2021}, and dark states \cite{Ozhigov2020}.
	
	In this paper, we discuss modifications of finite-dimensional QED models that allow us to interpret chemical reactions in terms of artificial atoms and molecules on quantum dots positioned inside optical cavities. Between the cavities, quantum motion of nuclei is permissible. Association reaction differ only in the initial states. By using the Lindblad operators of photon leakage from the cavity to the external environment to solve the single quantum master equation (QME), chemical processes with two-level atoms are schematically described. QME approach has been used to examine the dynamics of quantum open system \cite{Breuer2002}, and it is consistent with the principles of quantum thermodynamics \cite{Alicki1979, Kosloff2013}. Only Markovian approximations are applicable.
	
	This paper is organized as follows. After introducing the association--dissociation model of the neutral hydrogen molecule in Sec. \ref{sec:AssDissModel}, describing hybridization and de-hybridization of a couple of two-level artificial atoms, and based on the TCHM \cite{Angelakis2007}, we introduce electron spin-flip in Sec. \ref{sec:SpinTrans}. We also take into account how temperature changes in photonic modes affect quantum evolution and the formation of neutral hydrogen molecules in Sec. \ref{sec:Thermally}. In Sec. \ref{sec:BondPhonon}, a more accurate model incorporating a phonon and a covalent bond is raised. We offer a numerical technique to achieve complexity reduction in Sec. \ref{sec:Method}. We present the results of our numerical simulations in Sec. \ref{sec:Simulation}. Some brief comments on our results and extension to future work in Sec. \ref{sec:ConcluFuture} close out the paper. Some technical details are included in Appendices \ref{appx:ComExpTCH}, \ref{appx:Theorem} and \ref{appx:TensorGenerator}. List of abbreviations and notations used in this paper Tab. \ref{appx:Table} is put in Appendix \ref{appx:AbbreviationsNotations}.

	\section{The Association--dissociation model of neutral hydrogen molecule}
	\label{sec:AssDissModel}
	
	The formation of molecular hydrogen through a direct association of atoms has been studied mainly in connection with interstellar gas \cite{Latter1991, Wakelam2017}, where such an association is stimulated by photons emitted by stars. In this case, there is a large run of atoms before the collision, so that a semi classical description of the dynamics for the motion of atoms is possible; the statistics are determined by the Boltzmann distribution. In other works on the formation of molecular hydrogen, adsorption mechanisms (Eley--Rideal or Langmuir--Hinshelwood Mechanisms) on surfaces of the dust grains have been considered, which implies approximate calculation methods \cite{Cazaux2002}.
	
	Compared with these methods, we are investigating conditional hydrogen atoms (this may be a couple of other atoms that can be associated into a molecule), which move very slowly, so slowly that the characteristic action is comparable to Planck's constant, and it is impossible to apply even a semi-classical method, it is necessary to use a purely quantum type of description of dynamics. Such a process does not take place in empty space \cite{Latter1991}, but in a medium where the kinetic energy of atoms is extinguished by other atoms of the medium, so that their movements near the association point become purely quantum. So, our model is based on the first principles of quantum theory that permits its scaling to the large systems without the additional suppositions. However, the standard simplified representation of molecular orbital (MO) of hydrogen as a simple linear combination of atomic orbitals (AO) 1s is not suitable for describing the dynamics of association--dissociation of molecules, since this process in reality contains many intermediate states associated with the emission and absorption of photons, as well as the exchange of photons between two close atoms \cite{Jentschura2023}. The most significant intermediate state is associated with the formation of hybrid molecular orbitals and the emission of a photon during association, or with the decay of such orbitals during dissociation; such processes cannot be described by a simple hybridization of 1s orbitals. In addition, the atomic orbitals of the approaching atoms themselves differ from the stationary orbits of the free hydrogen atom, since the electron clouds are strongly deformed when the atoms approach. Therefore, we have supplemented the standard model with another intermediate state of electrons in atoms, which is obtained with such deformation. The state of a free atom is denoted by $|-1\rangle$, and $|0\rangle$ is the excited state of an electron in an atom, which is obtained when approaching another atom, so that the states $|0_1\rangle$ of the first and $|0_2\rangle$ of the second atom will hybridize into molecular orbitals.
	
	\begin{figure*}
		\centering
		\includegraphics[width=1.\textwidth]{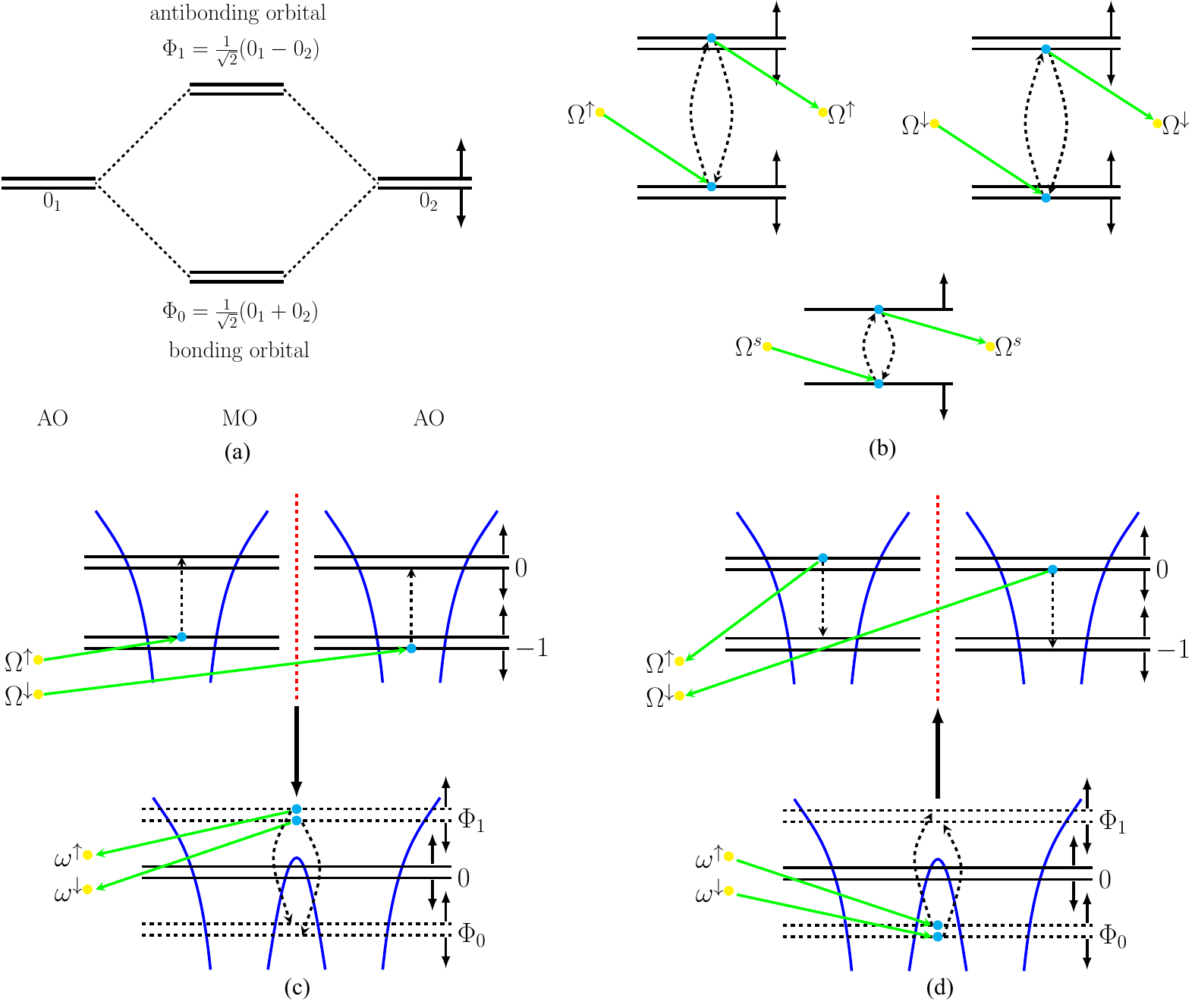}
		\caption{(online color) {\it The association-dissociation model of neutral hydrogen molecule.} The hybridization of two hydrogen atoms' orbitals, as well as bonding and antibonding orbitals, are shown in panel (a). Panel (b) shows three kinds of atom--field interactions (excitation and relaxation) corresponding to three different photonic modes: $\Omega^{\uparrow}$, $\Omega^{\downarrow}$ and $\Omega^s$. Each interaction can be considered as a separate JCM. The formation of H$_2$ caused by the association reaction of two hydrogen atoms is depicted in panel (c). The decomposition of H$_2$ caused by the dissociation reaction of these hydrogen atoms is depicted in panel (d). In the panels (b), (c) and (d), the blue and yellow dots, respectively, stand for electrons and photons.}
		\label{fig:AssDissModel}
	\end{figure*}
	
	The association--dissociation model of the neutral hydrogen molecule is modified from the TCHM (see Appendix \ref{appx:ComExpTCH}). In this model, each energy level, both atomic and molecular, is split into two levels with the same energy (approximately the same, with accuracy to Stark splitting): spin up and spin down, which are indicated by the signs $\uparrow$ and $\downarrow$, respectively. To differentiate each level on the spin, we will add these marks that indicate the energy level. Now the levels will be twice as much, and for each level there must be no more than one electron according to Pauli exclusion principle \cite{Pauli1925}. Thus, photons that excite the electron will be of the same type as the chosen spin direction.
	
	Hybridization of atomic orbitals and formation of molecular orbitals are shown in Fig. \ref{fig:AssDissModel}(a), where bonding orbital takes the form $\Phi_0=1/\sqrt{2}\left(0_1+0_2\right)$ and antibonding orbital takes the form $\Phi_1=1/\sqrt{2}\left(0_1-0_2\right)$. Interactions of atom with field are shown in panel (b) of Fig. \ref{fig:AssDissModel}, including excitation and relaxation of electron corresponding to $\Omega^{\uparrow}$, $\Omega^{\uparrow}$ and $\Omega^s$, respectively. In addition, photonic modes $\omega^{\uparrow}$ and $\omega^{\uparrow}$ also have the same above interactions as photonic modes $\Omega^{\uparrow}$ and $\Omega^{\uparrow}$. The electrons will be bound in the potential wells that each nucleus creates around itself. Fig. \ref{fig:AssDissModel}(c) displays the association reaction of H$_2$. Two electrons in the atomic ground orbital $-1$ with significant gaps between their nuclei, which correspond to two distinct spin directions, absorb respectively photons with modes $\Omega^{\uparrow}$ or $\Omega^{\downarrow}$, before rising to the atomic excited orbital $0$. At this time, the two excited state atoms can approach each other via quantum tunnelling effect and the potential barrier between the two potential wells decreases. Since the two electrons are in atomic excited orbitals, the atomic orbitals are hybridized into molecular excited orbitals, and the electrons are released on the molecular excited orbital $\Phi_1$. Then, two electrons quickly emit photons with the modes $\omega^{\uparrow}$ or $\omega^{\downarrow}$, respectively, and fall to the molecular ground orbital $\Phi_0$. Stable molecule is formed. Fig. \ref{fig:AssDissModel}(d) depicts the dissociation reaction of H$_2$. Two electrons in the molecular ground orbital absorb respectively photon with modes $\omega^{\uparrow}$ or $\omega^{\downarrow}$, rising to the molecular excited orbital as a result. The potential barrier rises, the molecular orbitals de-hybridized into atomic orbitals, and the electrons are liberated on the atomic excited orbital when nuclei scatter in various cavities. Finally, two electrons emit a photon with modes $\Omega^{\uparrow}$ or $\Omega^{\downarrow}$, and fall to the atomic ground orbital. The molecule disintegrates.
	
	In this paper we only consider two electrons with $\downarrow$ as the initial condition. We suppose that every type of photon has a sufficiently large wavelength to interact with an electron located in any cavity.

	The excited states of the electron with the spins for the first nucleus are indicated by $|0_1^{\uparrow}\rangle_e$ and $|0_1^{\downarrow}\rangle_e$. Usually simply written as $|0_1\rangle_e$, which can denote both $|0_1^{\uparrow}\rangle_e$ and $|0_1^{\downarrow}\rangle_e$. The first nucleus's ground electron states are then determined by $|-1_1\rangle_e$. For the second nucleus --- $|0_2\rangle_e$ and $|-1_2\rangle_e$. Only at great distances between nuclei are the ground states possible (see Figs. \ref{fig:AssDissModel}(c) and \ref{fig:AssDissModel}(d), where a vertical red dashed line indicates a significant distance between the nuclei). Possible only for atomic excited states $|0_{1,2}\rangle_e$ is orbital hybridization. Hybridization is impossible for the atomic ground states $|-1_{1,2}\rangle_e$. Hybrid molecular states of the electron energy are denoted by 
	\begin{subequations}
		\label{eq:MolState}
		\begin{align}
			&|\Phi_1\rangle_e=\frac{1}{\sqrt{2}}\left(|0_1\rangle_e-|0_2\rangle_e\right)\label{eq:MolStatePhi1}\\
			&|\Phi_0\rangle_e=\frac{1}{\sqrt{2}}\left(|0_1\rangle_e+|0_2\rangle_e\right)\label{eq:MolStatePhi0}
		\end{align}
	\end{subequations}
	where $|\Phi_1\rangle_e$ is molecular excited state, $|\Phi_0\rangle_e$ is molecular ground state.

	We introduce the second quantization, also known as the occupation number representation \cite{Dirac1927, Fock1932}, to prevent the difficulty that antisymmetrization causes from becoming more complicated. In this approach, the quantum many-body states are represented in the Fock state basis, which are constructed by filling up each single-particle state with a certain number of identical particles
	\begin{equation}
		\label{eq:FockState}
		|Fock\rangle=|n_1,n_2,n_3,\cdots,n_{\alpha},\cdots\rangle
	\end{equation}
	
	In the single-particle state $|\alpha\rangle$, it signifies that there are $n_{\alpha}$ particles. The total number of particles $N$ is equal to the sum of the occupation numbers, or $\sum_{\alpha}n_{\alpha}=N$. Due to the Pauli exclusion principle, the occupancy number $n_{\alpha}$ for fermions can only be $0$ or $1$ but it can be any non-negative integer for bosons. The many-body Hilbert space, also known as Fock space, is completely based on all of the Fock states. A linear collection of Fock states can be used to express any generic quantum many-body state. The creation and annihilation operators are introduced in the second quantization formalism to construct and handle the Fock states, giving researchers studying the quantum many-body theory useful tools.
	
	As a result, the entire system's Hilbert space for quantum states is $\mathcal{C}$ and takes the following form
	\begin{widetext}
		\begin{equation}
			\label{eq:SpaceC}
			|\Psi\rangle_{\mathcal{C}}=\underbrace{|p_1\rangle_{\omega^{\uparrow}}|p_2\rangle_{\omega^{\downarrow}}|p_3\rangle_{\Omega^{\uparrow}}|p_4\rangle_{\Omega^{\downarrow}}|p_5\rangle_{\Omega^s}}_{|photon\rangle}\underbrace{|l_1\rangle_{\substack{at_1\\or_0}}^{\uparrow}|l_2\rangle_{\substack{at_1\\or_0}}^{\downarrow}|l_3\rangle_{\substack{at_1\\or_{-1}}}^{\uparrow}|l_4\rangle_{\substack{at_1\\or_{-1}}}^{\downarrow}|l_5\rangle_{\substack{at_2\\or_0}}^{\uparrow}|l_6\rangle_{\substack{at_2\\or_0}}^{\downarrow}|l_7\rangle_{\substack{at_2\\or_{-1}}}^{\uparrow}|l_8\rangle_{\substack{at_2\\or_{-1}}}^{\downarrow}}_{|electron\rangle\ \mathrm{or}\ |orbital\rangle}\underbrace{|k\rangle_n}_{|nucleus\rangle}
		\end{equation}
	\end{widetext}
	where the quantum state consists of three parts: photon state $|photon\rangle$, electron state $|electron\rangle$ (or orbital state $|orbital\rangle$) and nucleus state $|nucleus\rangle$. The numbers of molecule photons with the modes $\omega^{\uparrow}$, $\omega^{\downarrow}$ are $p_1,\ p_2$, respectively; $p_3,\ p_4$ are the numbers of atomic photons with modes $\Omega^{\uparrow}$, $\Omega^{\downarrow}$, respectively; $p_5$ is the number of photons with mode $\Omega^s$, which can excite the electron spin from $\downarrow$ to $\uparrow$ in the atom. $l_{i,i\in\left\{1,2,\cdots,8\right\}}$ describes orbital state (each atom has four orbitals: $0^{\uparrow}$, $0^{\downarrow}$, $-1^{\uparrow}$ and $-1^{\downarrow}$): $l_i=1$ --- the orbital is occupied by one electron, $l_i=0$ --- the orbital is freed. The states of the nuclei are denoted by $|k\rangle_n$: $k=0$ --- state of nuclei, gathering together in one cavity, $k=1$ --- state of nuclei, scattering in different cavities.
	
	The space of quantum states $\mathcal{C}$ can be absolutely separated to two subspaces $\mathcal{A}$ and $\mathcal{D}$, where $\mathcal{A}\oplus\mathcal{D}=\mathcal{C}$, $\mathcal{A}\cap\mathcal{D}=\vec{0}$. The subspace for associative system, also known as molecular system, in which states correspond to $|0\rangle_n$, is $\mathcal{A}$. The subspace for dissociative system, also known as atomic system, in which the states correspond to $|1\rangle_n$, is $\mathcal{D}$. The following are the definitions for $\mathcal{A}$ and $\mathcal{D}$
	\begin{widetext}
		\begin{subequations}
			\label{eq:Subspaces}
			\begin{align}
				\mathcal{A}&=\sum_{\substack{p_{i_1,i_1\in\left\{1,2,\cdots,5\right\}},\\l_{i_2,i_2\in\left\{1,2,\cdots,8\right\}}}}c^0_{p_{i_1},l_{i_2}}\underbrace{|p_1\rangle_{\omega^{\uparrow}}|p_2\rangle_{\omega^{\downarrow}}\cdots|p_5\rangle_{\Omega^s}}_{|photon\rangle}\underbrace{|l_1\rangle_{\substack{at_1\\or_0}}^{\uparrow}|l_2\rangle_{\substack{at_1\\or_0}}^{\downarrow}\cdots|l_8\rangle_{\substack{at_2\\or_{-1}}}^{\downarrow}}_{|electron\rangle}\underbrace{|0\rangle_n}_{|nucleus\rangle}\label{eq:SubspaceA}\\
				\mathcal{D}&=\sum_{\substack{p_{i_1,i_1\in\left\{1,2,\cdots,5\right\}},\\l_{i_2,i_2\in\left\{1,2,\cdots,8\right\}}}}c^1_{p_{i_1},l_{i_2}}\underbrace{|p_1\rangle_{\omega^{\uparrow}}|p_2\rangle_{\omega^{\downarrow}}\cdots|p_5\rangle_{\Omega^s}}_{|photon\rangle}\underbrace{|l_1\rangle_{\substack{at_1\\or_0}}^{\uparrow}|l_2\rangle_{\substack{at_1\\or_0}}^{\downarrow}\cdots|l_8\rangle_{\substack{at_2\\or_{-1}}}^{\downarrow}}_{|electron\rangle}\underbrace{|1\rangle_n}_{|nucleus\rangle}\label{eq:SubspaceD}
			\end{align}
		\end{subequations}
	\end{widetext}
	where $c^0_{p_{i_1},l_{i_2}},\ c^1_{p_{i_1},l_{i_2}}$ are normalization factors.
	
	The association--dissociation model of the neutral hydrogen molecule used in this paper is an adaptation of the TCHM that incorporates a multi-mode electromagnetic field inside optical cavities. The standard TCM describes the interaction of $N$ two-level atoms with a single-mode electromagnetic field inside an optical cavity and has been generalized to several cavities coupled by an optical fibre --- the standard TCHM. First, the dynamics of system is described by solving the QME for the density matrix with the Lindblad operators of photon leakage from the cavity to external environment. The QME in the Markovian approximation for the density operator $\rho$ of the system takes the following form
	\begin{equation}
		\label{eq:QME}
		i\hbar\dot{\rho}=\mathcal{L}\left(\rho\right)=\left[H,\rho\right]+iL\left(\rho\right)
	\end{equation}
	where $\mathcal{L}\left(\rho\right)$ is Lindblad superoperator and $\left[H,\rho\right]=H\rho-\rho H$ is the commutator. We have a graph $\mathcal{K}$ of the potential photon dissipations between the states that are permitted. The edges and vertices of $\mathcal{K}$ represent the permitted dissipations and the states, respectively. Similar to this, $\mathcal{K}'$ is a graph of potential photon influxes that are permitted. $L\left(\rho\right)$ is as follows
	\begin{equation}
		\label{eq:LindbladOperator}
		L\left(\rho\right)=\sum_{k\in \mathcal{K}} L_k\left(\rho\right)+\sum_{k'\in \mathcal{K}'} L_{k'}\left(\rho\right)
	\end{equation}
	where $L_k\left(p\right)$ is the standard dissipation superoperator corresponding to the jump operator $A_k$ and taking as an argument on the density matrix $\rho$
	\begin{equation}
		\label{eq:DissSuper}
		L_k\left(\rho\right)=\gamma_k\left(A_k\rho A_k^{\dag}-\frac{1}{2}\left\{\rho, A_k^{\dag}A_k\right\}\right)
	\end{equation}
	where $\left\{\rho, A_k^{\dag}A_k\right\}=\rho A_k^{\dag}A_k + A_k^{\dag}A_k\rho$ is the anticommutator. 
	The term $\gamma_k$ refers to the overall spontaneous emission rate for photons for $k\in \mathcal{K}$ caused by photon leakage from the cavity to the external environment. Similarly, $L_{k'}\left(p\right)$ is the standard influx superoperator, having the following form
	\begin{equation}
		\label{eq:InfluxSuper}
		L_{k'}\left(\rho\right)=\gamma_{k'}\left(A_k^{\dag}\rho A_k-\frac{1}{2}\left\{\rho, A_kA_k^{\dag}\right\}\right)
	\end{equation}
	The total spontaneous influx rate for photon for $k'\in \mathcal{K}'$ is denoted by $\gamma_{k'}$. 

	The coupled-system Hamiltonian of the association--dissociation model in Eq. \eqref{eq:QME} is expressed by the total energy operator 
	\begin{equation}
		\label{eq:Hamil}
		H=H_{\mathcal{A}}+H_{\mathcal{D}}+H_{tun}
	\end{equation}
	where $H_{tun}$ denotes the quantum tunnelling effect between $H_{\mathcal{A}}$ and $H_{\mathcal{D}}$, which are the associative and dissociative Hamiltonians, respectively, that correspond to $\mathcal{A}$ and $\mathcal{D}$.
	
	$H_{\mathcal{A}}$ has following form
	\begin{equation}
		\label{eq:HamilA}
		H_{\mathcal{A}}=\left(H_{\mathcal{A},field}+H_{\mathcal{A},mol}+H_{\mathcal{A},int}\right)\sigma_n\sigma_n^{\dag}
	\end{equation}
	where $\sigma_n\sigma_n^{\dag}$ verifies that nuclei are close.
	
	Rotating wave approximation (RWA) is taken into account. This approach ignores the quickly oscillating terms $\sigma^{\dag}a^{\dag},\ \sigma a$ in a Hamiltonian. When the strength of the applied electromagnetic radiation is close to resonance with an atomic transition and the intensity is low, this approximation holds true \cite{Wu2007}. Thus,
	\begin{equation}
		\label{eq:RWACondition}
		\frac{g}{\hbar\omega_c}\approx\frac{g}{\hbar\omega_n}\ll 1
	\end{equation}
where $\omega_c$ stands for cavity frequency; and $\omega_n$ for transition frequency, which includes $\omega$ ($\omega^{\uparrow}$ and $\omega^{\downarrow}$) for molecule, and $\Omega$ ($\Omega^{\uparrow}$ and $\Omega^{\downarrow}$) for atom. RWA allows us to change $\left(\sigma^{\dag}+\sigma\right)\left(a^{\dag}+a\right)$ to $\sigma^{\dag}a+\sigma a^{\dag}$ in Eqs. \eqref{eq:HamilAInt} and \eqref{eq:HamilDInt}. We typically presume that $\omega_c=\omega_n$. Thus,
	\begin{subequations}
		\label{eq:HamilADetail}
		\begin{align}
			&H_{\mathcal{A},field}=\hbar\omega^{\uparrow}a_{\omega^{\uparrow}}^{\dag}a_{\omega^{\uparrow}}+\hbar\omega^{\downarrow}a_{\omega^{\downarrow}}^{\dag}a_{\omega^{\downarrow}}\label{eq:HamilAField}\\
			&H_{\mathcal{A},mol}=\hbar\omega^{\uparrow}\sigma_{\omega^{\uparrow}}^{\dag}\sigma_{\omega^{\uparrow}}+\hbar\omega^{\downarrow}\sigma_{\omega^{\downarrow}}^{\dag}\sigma_{\omega^{\downarrow}}\label{eq:HamilAMol}\\
			&H_{\mathcal{A},int}=g_{\omega^{\uparrow}}\left(a_{\omega^{\uparrow}}^{\dag}\sigma_{\omega^{\uparrow}}+a_{\omega^{\uparrow}}\sigma_{\omega^{\uparrow}}^{\dag}\right)\\
			&+g_{\omega^{\downarrow}}\left(a_{\omega^{\downarrow}}^{\dag}\sigma_{\omega^{\downarrow}}+a_{\omega^{\downarrow}}\sigma_{\omega^{\downarrow}}^{\dag}\right)\label{eq:HamilAInt}
		\end{align}
	\end{subequations}
	where $\hbar=h/2\pi$ is the reduced Planck constant or Dirac constant. $H_{\mathcal{A},field}$ is the photon energy operator, $H_{\mathcal{A},mol}$ is the molecule energy operator, $H_{\mathcal{A},int}$ is the molecule--photon interaction operator. $g_{\omega}$ is the coupling strength between the photon mode $\omega$ (with annihilation and creation operators $a_{\omega}$ and $a_{\omega}^{\dag}$, respectively) and the electrons in the molecule (with excitation and relaxation operators $\sigma_{\omega}^{\dag}$ and $\sigma_{\omega}$, respectively).

	Then $H_{\mathcal{D}}$ is described in following form
	\begin{equation}
		\label{eq:HamilD}
		H_{\mathcal{D}}=\left(H_{\mathcal{D},field}+H_{\mathcal{D},mol}+H_{\mathcal{D},int}\right)\sigma_n^{\dag}\sigma_n
	\end{equation}
where $\sigma_n^{\dag}\sigma_n$ verifies that nuclei are far away. Similarly, we introduce RWA
	\begin{subequations}
		\label{eq:HamilDDetail}
		\begin{align}
			&H_{\mathcal{D},field}=\hbar\Omega^{\uparrow}a_{\Omega^{\uparrow}}^{\dag}a_{\Omega^{\uparrow}}+\hbar\Omega^{\downarrow}a_{\Omega^{\downarrow}}^{\dag}a_{\Omega^{\downarrow}}\label{eq:HamilDField}\\
			&H_{\mathcal{D},at}=\sum_{i=1,2}\left(\hbar\Omega^{\uparrow}\sigma_{\Omega^{\uparrow},i}^{\dag}\sigma_{\Omega^{\uparrow},i}+\hbar\Omega^{\downarrow}\sigma_{\Omega^{\downarrow},i}^{\dag}\sigma_{\Omega^{\downarrow},i}\right)\label{eq:HamilDAt}\\
			&H_{\mathcal{D},int}=\sum_{i=1,2}\left\{g_{\Omega^{\uparrow}}\left(a_{\Omega^{\uparrow}}^{\dag}\sigma_{\Omega^{\uparrow},i}+a_{\Omega^{\uparrow}}\sigma_{\Omega^{\uparrow},i}^{\dag}\right)\right.\\
			&\left.+g_{\Omega^{\downarrow}}\left(a_{\Omega^{\downarrow}}^{\dag}\sigma_{\Omega^{\downarrow},i}+a_{\Omega^{\downarrow}}\sigma_{\Omega^{\downarrow},i}^{\dag}\right)\right\}\label{eq:HamilDInt}
		\end{align}
	\end{subequations}
where $H_{\mathcal{D},field}$ is the photon energy operator, $H_{\mathcal{D},at}$ is the atom energy operator, $H_{\mathcal{D},int}$ is atom--photon interaction operator. $g_{\Omega}$ is the coupling strength between the photon mode $\Omega$ (with annihilation and creation operators $a_{\Omega}$ and $a_{\Omega}^{\dag}$, respectively) and the electrons in the atom (with excitation and relaxation operators $\sigma_{\Omega,i}^{\dag}$ and $\sigma_{\Omega,i}$, respectively, here $i$ denotes index of atoms).

	\begin{figure*}
		\centering
		\includegraphics[width=1\textwidth]{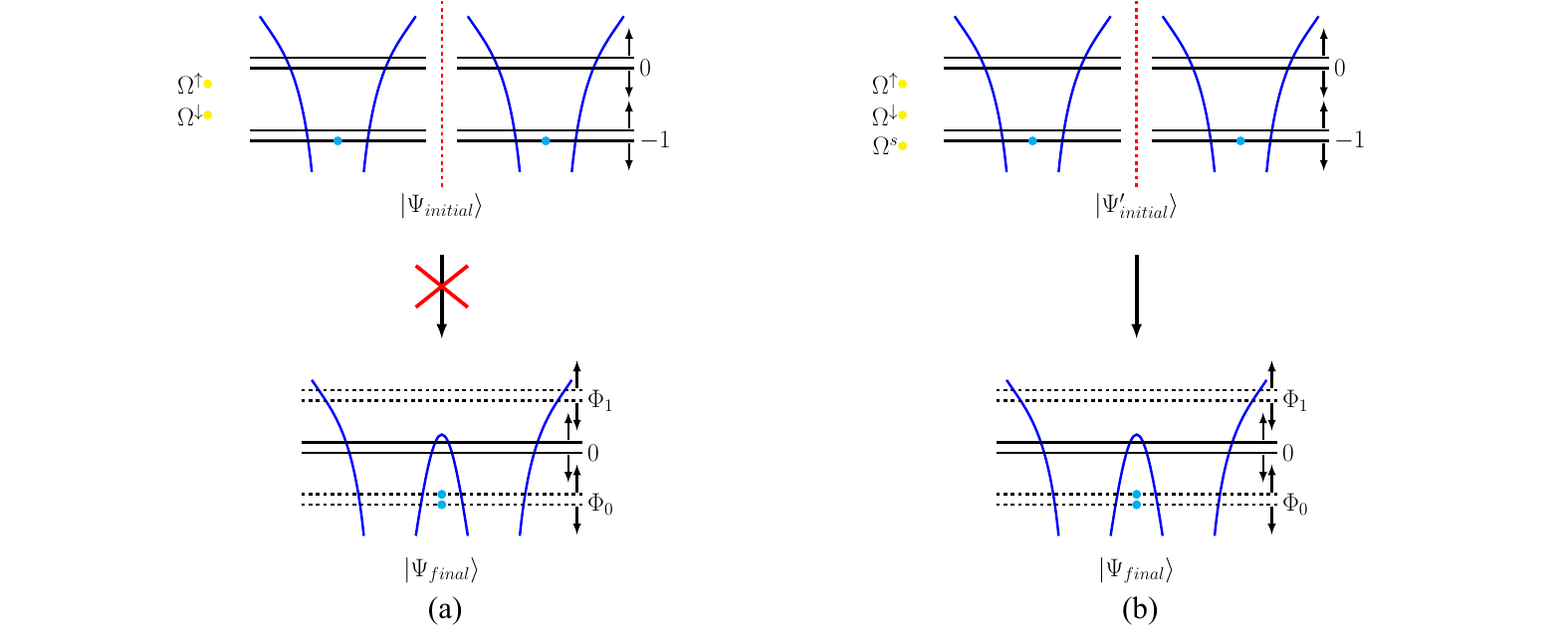}
		\caption{(online color) {\it Electron spin transition.} The situation without consideration of electron spin transition is depicted in panel (a), where it is impossible to construct a neutral hydrogen molecule if only two photons with the same modes $\Omega^{\downarrow}$ are present at the beginning. The situation, which takes into account the electron spin transition, is depicted in panel (b), where the addition of a photon with the mode $\Omega^s$ can result in the formation of a neutral hydrogen molecule.}
		\label{fig:SpinTrans}
	\end{figure*}

	Finally, $H_{tun}$ describe the hybridization and de-hybridization, realized by quantum tunnelling effect, it takes the form
	\begin{equation}
		\label{eq:HamilTDetail}
		\begin{aligned}
			H_{tun}&=\zeta_2\sigma_{\omega^{\uparrow}}^{\dag}\sigma_{\omega^{\uparrow}}\sigma_{\omega^{\downarrow}}^{\dag}\sigma_{\omega^{\downarrow}}\left(\sigma_n^{\dag}+\sigma_n\right)\\
			&+\zeta_1\sigma_{\omega^{\uparrow}}\sigma_{\omega^{\uparrow}}^{\dag}\sigma_{\omega^{\downarrow}}^{\dag}\sigma_{\omega^{\downarrow}}\left(\sigma_n^{\dag}+\sigma_n\right)\\
			&+\zeta_1\sigma_{\omega^{\uparrow}}^{\dag}\sigma_{\omega^{\uparrow}}\sigma_{\omega^{\downarrow}}\sigma_{\omega^{\downarrow}}^{\dag}\left(\sigma_n^{\dag}+\sigma_n\right)\\
			&+\zeta_0\sigma_{\omega^{\uparrow}}\sigma_{\omega^{\uparrow}}^{\dag}\sigma_{\omega^{\downarrow}}\sigma_{\omega^{\downarrow}}^{\dag}\left(\sigma_n^{\dag}+\sigma_n\right)
		\end{aligned}
	\end{equation}
	where $\sigma_{\omega^{\uparrow}}^{\dag}\sigma_{\omega^{\uparrow}}\sigma_{\omega^{\downarrow}}^{\dag}\sigma_{\omega^{\downarrow}}$ verifies that two electrons with different spins are at orbital $\Phi_1$ with large tunnelling intensity $\zeta_2$; $\sigma_{\omega^{\uparrow}}\sigma_{\omega^{\uparrow}}^{\dag}\sigma_{\omega^{\downarrow}}^{\dag}\sigma_{\omega^{\downarrow}}$ verifies that electron with $\uparrow$ is at orbital $\Phi_0$ and electron with $\downarrow$ is at orbital $\Phi_1$, with low tunnelling intensity $\zeta_1$; $\sigma_{\omega^{\uparrow}}^{\dag}\sigma_{\omega^{\uparrow}}\sigma_{\omega^{\downarrow}}\sigma_{\omega^{\downarrow}}^{\dag}$ verifies that electron with $\uparrow$ is at orbital $\Phi_1$ and electron with $\downarrow$ is at orbital $\Phi_0$, with low tunnelling intensity $\zeta_1$; $\sigma_{\omega^{\uparrow}}\sigma_{\omega^{\uparrow}}^{\dag}\sigma_{\omega^{\downarrow}}\sigma_{\omega^{\downarrow}}^{\dag}$ verifies that two electrons with different spins are at orbital $\Phi_0$ with tunnelling intensity $\zeta_0$, which equal to $0$. In a nutshell, the quantum tunnelling effect is diminished when an electron fall to the molecular ground state.

	\section{Electron spin transition} 
	\label{sec:SpinTrans}
	
	The association--dissociation model is introduced with spin photons with mode $\Omega^s$ in this section, allowing for transitions between $\uparrow$ and $\downarrow$. The Pauli exclusion principle, which prohibits the presence of electrons with the same spin at the same energy level, must be carefully followed by electron spins. We agree that an electron spin transition is only possible if the electrons are in the atomic states corresponding to $|1\rangle_n$. Electron spin transition is forbidden when electrons are in molecular states corresponding to $|0\rangle_n$, which contravenes Pauli exclusion principle. Only a state with two electrons in orbital $\Phi_0$ with different spins can result in the stable formation of H$_2$. This situation is just right accord with that a stable system has a lower energy level, this position is ideal.
	
	The Hamiltonian of electron spin transition takes the form
	\begin{equation}
		\label{eq:HamilSpin}
		\begin{aligned}
			H_{spin}&=\hbar\Omega^sa_{\Omega^s}^{\dag}a_{\Omega^s}+\hbar\Omega^s\sum_{i=1,2}\sigma_{\Omega^s,i}^{\dag}\sigma_{\Omega^s,i}\\
			&+g_{\Omega^s}\sum_{i=1,2}\left(a_{\Omega^s}^{\dag}\sigma_{\Omega^s,i}+a_{\Omega^s}\sigma_{\Omega^s,i}^{\dag}\right)
		\end{aligned}
	\end{equation}
	where $i$ denotes index of atoms. And total Hamiltonian can be rewritten as follows
	\begin{equation}
		\label{eq:NewHamil}
		H=H_{\mathcal{A}}+H_{\mathcal{D}}+H_{tun}+H_{spin}
	\end{equation}	 
	
	We consider two situations:
	\begin{itemize}
		\item in Fig. \ref{fig:SpinTrans}(a) we only pump into two photons with different modes $\Omega^{\uparrow}$ and $\Omega^{\downarrow}$, and spin photons are proviso not taken into consideration, and transition between $\uparrow$ and $\downarrow$ is prohibited;
		\item in Fig. \ref{fig:SpinTrans}(b) spin photons and corresponding transition is introduced.
	\end{itemize}
	Theoretically, the formation of H$_2$ is thus impossible in the first situation, and is achieved in the second situation. 

	\section{Thermally stationary state} 
	\label{sec:Thermally}
		
	As a mixed state with a Gibbs distribution of Fock components, we define the stationary state of a field with temperature $T$ as follows
	\begin{equation}
		\label{eq:Photon_gibbs}
		{\cal G}\left(T\right)_f=c \sum\limits_{p=0}^\infty exp\left(-\frac{\hbar\omega_c p}{KT}\right)|p\rangle\langle p|
	\end{equation} 
where $K$ is the Boltzmann constant, $c$ is the normalization factor, $p$ is the number of photons, $\omega_c$ is the photonic mode. The notation $\gamma_{k'}/\gamma_{k}=\mu$ is presented. Since the temperature would otherwise be endlessly high and the state ${\cal G}\left(T\right)_f$ would not be normalizable, the state will then only exist at $\mu<1$.

	The probability of the photonic Fock state $|p\rangle$ at temperature $T$ is proportional to $exp\left(-\frac{\hbar\omega_c}{KT}\right)$. In our model, we assume
	\begin{equation}
		\label{eq:PopulationFock}
		\mu=exp\left(-\frac{\hbar\omega_c}{KT}\right)
	\end{equation}
	from where $T=\frac{\hbar\omega_c}{K\ln\left(1/\mu\right)}$.

	The following theorem takes place as follows \cite{Kulagin2018} and the proof of it is given in the Appendix \ref{appx:Theorem}:

	The thermally stationary state of atoms and fields at temperature $T$ has the form $\rho_{state}=\rho_{ph}\otimes\rho_{at}$, where $\rho_{ph}$ is the state of the photon and $\rho_{at}$ is the state of the atom.
	
	\begin{figure*}
		\centering
		\includegraphics[width=1.\textwidth]{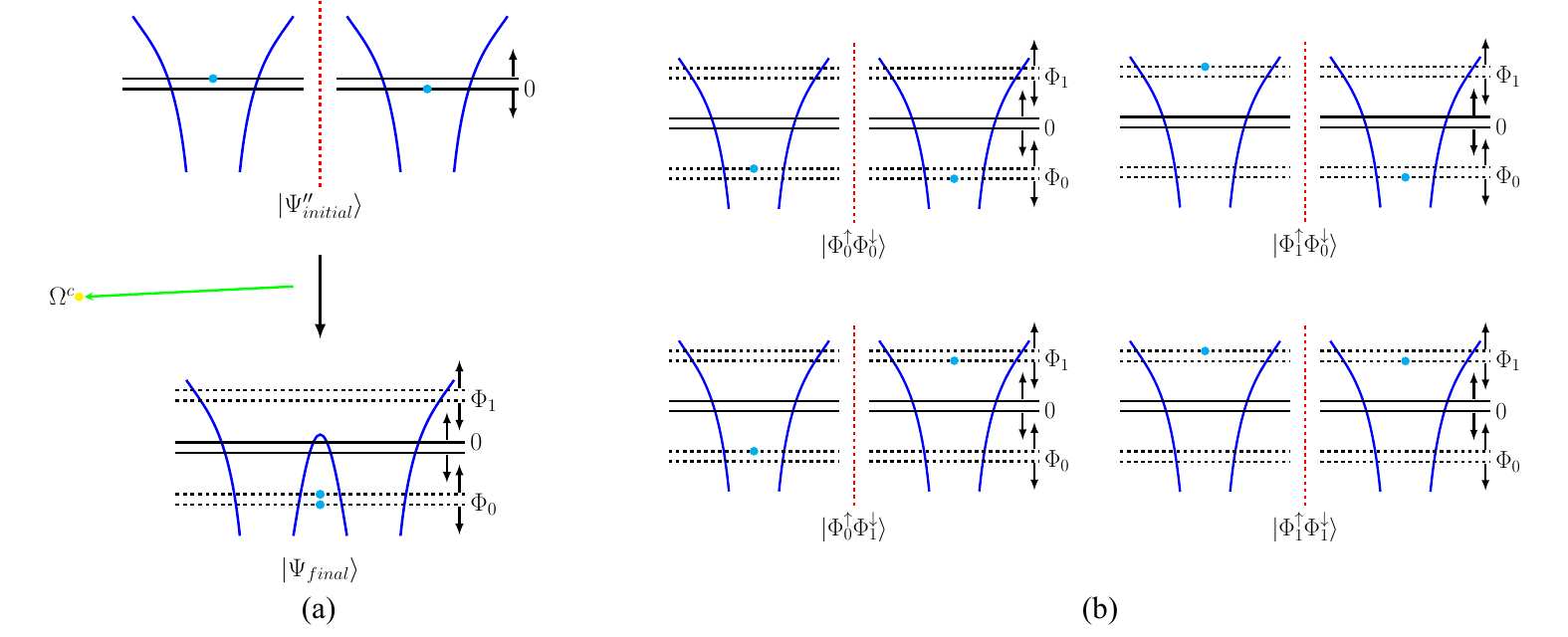}
		\caption{(online color) {\it The model with covalent bond and phonon.} When nuclei are in distinct cavities in the model depicted in panel (a), electrons are constrained to orbital $0$; however, when nuclei are in the same cavity, electrons can jump between orbitals $\Phi_1$ and $\Phi_0$. In order to make a covalent bond, a phonon must be released, and in order to break a covalent link, a phonon must be absorbed. In panel (b), according to Eqs. \eqref{eq:MolState}, $|\Psi_{initial}''\rangle$ can be decomposed into the sum of four states $|\Phi_0^{\uparrow}\Phi_0^{\downarrow}\rangle$, $|\Phi_1^{\uparrow}\Phi_0^{\downarrow}\rangle$, $|\Phi_0^{\uparrow}\Phi_1^{\downarrow}\rangle$ and $|\Phi_1^{\uparrow}\Phi_1^{\downarrow}\rangle$.}
		\label{fig:BondPhonon}
	\end{figure*}
	
	\section{The model with covalent bond and phonon} 
	\label{sec:BondPhonon}
	
	Now, a simpler and more precise model featuring a covalent bond and a simple harmonic oscillator (phonon) is presented in Fig. \ref{fig:BondPhonon}. Both the association reaction and the dissociation reaction can be interpreted by this model. The dissociation reaction cannot be fully explained by the prior model.
	
	The Hilbert space of quantum states of the entire system, having the following form
	\begin{equation}
		\label{eq:SpaceBondPhonon}
		|\Psi\rangle_{\mathcal{C}}=|p_1\rangle_{\omega^{\uparrow}}|p_2\rangle_{\omega^{\downarrow}}|m\rangle_{\Omega^c}|l_1\rangle_{\Phi_1^{\uparrow}}|l_2\rangle_{\Phi_1^{\downarrow}}|L\rangle_{cb}|k\rangle_{n}
	\end{equation}
	where $p_1,\ p_2$ are the numbers of molecular photons with modes $\omega^{\uparrow}$, $\omega^{\downarrow}$, respectively; $m$ is the number of phonons with mode $\Omega^c$. $l_1,\ l_2$ describe orbital state: $l_1=1$ --- electron with spin $\uparrow$ in excited orbital $\Phi_1^{\uparrow}$, $l_1=0$ --- electron with spin $\uparrow$ in ground orbital $\Phi_0^{\uparrow}$; $l_2=1$ --- electron with spin $\downarrow$ in excited orbital $\Phi_1^{\downarrow}$, $l_2=0$ --- electron with spin $\downarrow$ in ground orbital $\Phi_0^{\downarrow}$. The states of the covalent bond are denoted by $|L\rangle_{cb}$: $L=0$ --- covalent bond formation, $L=1$ --- covalent bond breaking. The states of the nuclei are denoted by $|k\rangle_n$: $k=0$ --- state of nuclei, gathering together in one cavity, $k=1$ --- state of nuclei, scattering in different cavities.
	
	Hamiltonian of this new model has following form
	\begin{equation}
		\label{eq:HamilBondPhonon}
		\begin{aligned}
			H_{cb}&=\hbar\omega^{\uparrow}a_{\omega^{\uparrow}}^{\dag}a_{\omega^{\uparrow}}+\hbar\omega^{\downarrow}a_{\omega^{\downarrow}}^{\dag}a_{\omega^{\downarrow}}+\hbar\Omega^c a_{\Omega^c}^{\dag}a_{\Omega^c}\\
			&+\hbar\omega^{\uparrow}\sigma_{\omega^{\uparrow}}^{\dag}\sigma_{\omega^{\uparrow}}+\hbar\omega^{\downarrow}\sigma_{\omega^{\downarrow}}^{\dag}\sigma_{\omega^{\downarrow}}+\hbar\Omega^c\sigma_{\Omega^c}^{\dag}\sigma_{\Omega^c}\\
			&+g_{\omega^{\uparrow}}\left(a_{\omega^{\uparrow}}^{\dag}\sigma_{\omega^{\uparrow}}+a_{\omega^{\uparrow}}\sigma_{\omega^{\uparrow}}^{\dag}\right)\sigma_{\Omega^c}\sigma_{\Omega^c}^{\dag}\\
			&+g_{\omega^{\downarrow}}\left(a_{\omega^{\downarrow}}^{\dag}\sigma_{\omega^{\downarrow}}+a_{\omega^{\downarrow}}\sigma_{\omega^{\downarrow}}^{\dag}\right)\sigma_{\Omega^c}\sigma_{\Omega^c}^{\dag}\\
			&+g_{\Omega^c}\left(a_{\Omega^c}^{\dag}\sigma_{\Omega^c}+a_{\Omega^c}\sigma_{\Omega^c}^{\dag}\right)\\
			&+\zeta\left(\sigma_n^{\dag}\sigma_n+\sigma_n\sigma_n^{\dag}\right)
		\end{aligned}
	\end{equation}
	where $\sigma_{\Omega^c}\sigma_{\Omega^c}^{\dag}$ verifies that covalent bond is formed. $g_{\Omega^c}$ --- strength of formation or breaking of covalent bond, $\zeta$ --- tunnelling intensity.
	
	Initial state $|\Psi_{initial}''\rangle$ is shown in Fig. \ref{fig:BondPhonon}(a), which can be decomposed into the sum of four states
	\begin{equation}
		\label{eq:InitialDecomposed}
		|\Psi_{initial}''\rangle=\frac{1}{2}\left(|\Phi_0^{\uparrow}\Phi_0^{\downarrow}\rangle+|\Phi_1^{\uparrow}\Phi_0^{\downarrow}\rangle-|\Phi_0^{\uparrow}\Phi_1^{\downarrow}\rangle-|\Phi_1^{\uparrow}\Phi_1^{\downarrow}\rangle\right)
	\end{equation}
	where
	\begin{subequations}
		\label{eq:PhiPhi}
		\begin{align}
			|\Phi_0^{\uparrow}\Phi_0^{\downarrow}\rangle=|0\rangle_{\omega^{\uparrow}}|0\rangle_{\omega^{\downarrow}}|0\rangle_{\Omega^c}|0\rangle_{\Phi_1^{\uparrow}}|0\rangle_{\Phi_1^{\downarrow}}|1\rangle_{cb}|1\rangle_{n}\label{eq:Phi0Phi0}\\
			|\Phi_1^{\uparrow}\Phi_0^{\downarrow}\rangle=|0\rangle_{\omega^{\uparrow}}|0\rangle_{\omega^{\downarrow}}|0\rangle_{\Omega^c}|1\rangle_{\Phi_1^{\uparrow}}|0\rangle_{\Phi_1^{\downarrow}}|1\rangle_{cb}|1\rangle_{n}\label{eq:Phi1Phi0}\\
			|\Phi_0^{\uparrow}\Phi_1^{\downarrow}\rangle=|0\rangle_{\omega^{\uparrow}}|0\rangle_{\omega^{\downarrow}}|0\rangle_{\Omega^c}|0\rangle_{\Phi_1^{\uparrow}}|1\rangle_{\Phi_1^{\downarrow}}|1\rangle_{cb}|1\rangle_{n}\label{eq:Phi0Phi1}\\
			|\Phi_1^{\uparrow}\Phi_1^{\downarrow}\rangle=|0\rangle_{\omega^{\uparrow}}|0\rangle_{\omega^{\downarrow}}|0\rangle_{\Omega^c}|1\rangle_{\Phi_1^{\uparrow}}|1\rangle_{\Phi_1^{\downarrow}}|1\rangle_{cb}|1\rangle_{n}\label{eq:Phi1Phi1}
		\end{align}
	\end{subequations}
	It should be noted that $|\Phi_0^{\uparrow}\Phi_0^{\downarrow}\rangle$ shown in Fig. \ref{fig:BondPhonon}(b) does not mean that there is the electron with $\uparrow$ in the ground state of the atom on the left, and the electron with $\downarrow$ in the ground state of the atom on the right. The exact reverse can be true. We only know that one of the atoms (we do not know which one) has the electron with $\uparrow$ in the ground state, and that the other atom has the electron with $\downarrow$. This is because we employ second quantization. In Fig. \ref{fig:BondPhonon}(b), for the convenience of explanation, we just intentionally fixed the electron with $\uparrow$ on the left atom. The same is true for the other three states $|\Phi_1^{\uparrow}\Phi_0^{\downarrow}\rangle$, $|\Phi_0^{\uparrow}\Phi_1^{\downarrow}\rangle$ and $|\Phi_1^{\uparrow}\Phi_1^{\downarrow}\rangle$.
	
	Now we define $\left\{0\succ_{cb}\right.$ and $\left\{1\succ_{cb}\right.$, which have following forms
	\begin{widetext}
		\begin{subequations}
			\label{eq:StateCBs}
			\begin{align}
				\left\{0\succ_{cb}\right.&=\sum_{p_1,p_2,m,l_1,l_2,k}c^2_{p_1,p_2,m,l_1,l_2,k}|p_1\rangle_{\omega^{\uparrow}}|p_2\rangle_{\omega^{\downarrow}}|m\rangle_{\Omega^c}|l_1\rangle_{\Phi_1^{\uparrow}}|l_2\rangle_{\Phi_1^{\downarrow}}|0\rangle_{cb}|k\rangle_n\label{eq:StateCB0}\\
				\left\{1\succ_{cb}\right.&=\sum_{p_1,p_2,m,l_1,l_2,k}c^3_{p_1,p_2,m,l_1,l_2,k}|p_1\rangle_{\omega^{\uparrow}}|p_2\rangle_{\omega^{\downarrow}}|m\rangle_{\Omega^c}|l_1\rangle_{\Phi_1^{\uparrow}}|l_2\rangle_{\Phi_1^{\downarrow}}|1\rangle_{cb}|k\rangle_n\label{eq:StateCB1}
			\end{align}
		\end{subequations}
	\end{widetext}
	where $c^2_{p_1,p_2,m,l_1,l_2,k},\ c^3_{p_1,p_2,m,l_1,l_2,k}$ are normalization factors.

	\section{Numerical method} 
	\label{sec:Method}

	The solution $\rho\left(t\right)$ in Eq. \eqref{eq:QME} may be approximately found as a sequence of two steps: in the first step we make one step in the solution of the unitary part of Eq. \eqref{eq:QME}
	\begin{equation}
		\label{eq:UnitaryPart}
		\tilde{\rho}\left(t+dt\right)=exp\left({-\frac{i}{\hbar}Hdt}\right)\rho\left(t\right)exp\left(\frac{i}{\hbar}Hdt\right)
	\end{equation}
and in the second step, make one step in the solution of Eq. \eqref{eq:QME} with the commutator removed:
	\begin{equation}
		\label{eq:Solution}
		\rho\left(t+dt\right)=\tilde{\rho}\left(t+dt\right)+\frac{1}{\hbar}L\left(\tilde{\rho}(t+dt)\right)dt
	\end{equation}
	
	The main problem of quantum many-body physics is the fact that the Hilbert space grows exponentially with size, which we call the curse of dimensionality. In order to solve this problem, several schemes including the density matrix renormalization group (DMRG) method \cite{White1992, White1993} have been proposed. Our task is to describe a qualitative scenario of chemical dynamics, so we take the following method.
	
	We have the conventional technique known as tensor product for establishing Hamiltonian in Eq. \eqref{eq:UnitaryPart}. Through the use of the tensor product, we can directly establish the Hamiltonian with Eq. \eqref{eq:Hamil}; however, the dimension of the Hamiltonian that results from this method is frequently very large and contains a lot of excess states that are not involved in evolution, particularly when the degree of freedom of the system is high. In this section, we will introduce the generator algorithm (comparison between tensor product and generator algorithm is shown in Appx. \ref{appx:TensorGenerator}), which is based on the occupation number representation in Eq. \eqref{eq:SpaceC}, and includes the following two steps:
	\begin{itemize}
		\item generating and numbering potential evolution states involved in the evolution in accordance with the initial state and any its potential dissipative states that may be relevant in solving QME;
		\item establishing Hamiltonian with these states and potential interactions and dissipations among them.
	\end{itemize}
	Using this technique, we now eliminate the extra unnecessary states and obtain anew $\mathcal{C}'$ and $H'$, where $\mathcal{C}'\subset\mathcal{C}$ and $dim\left(H'\right)\leq dim\left(H\right)$. In this paper, the $dim\left(H'\right)\approx 100$ is far smaller than the $dim\left(H\right)=2^{14}=16384$. As a result, complexity is reduced. The effectiveness of this reduction strategy increases with the increase of degree of freedom for multi-particle systems.

	\section{Simulations and results} 
	\label{sec:Simulation}
	
	\begin{figure*}
		\centering
		\includegraphics[width=1\textwidth]{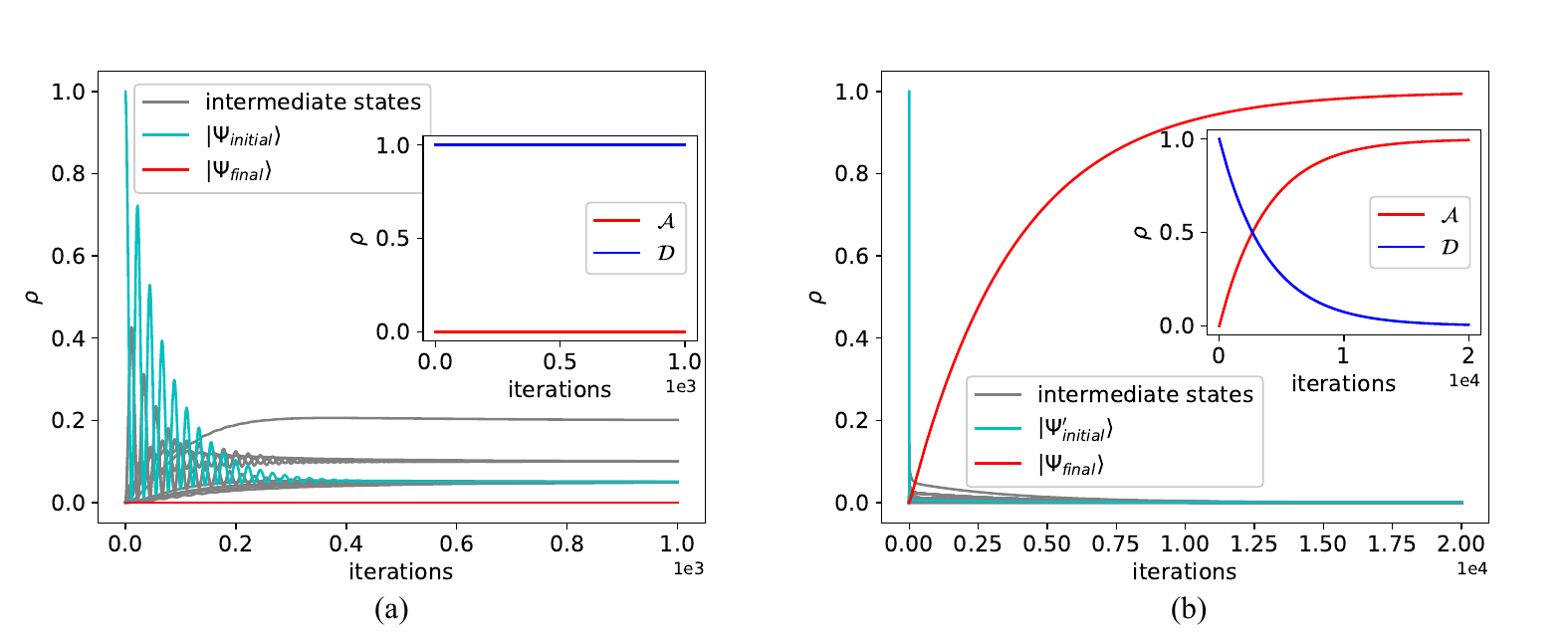}
		\caption{(online color) {\it The evolution without/with consideration of electron spin transition.} Panel (a) shows the evolution without consideration of electron spin transition. Probability of state $|\Psi_{initial}\rangle$ is denoted by cyan solid curve, probability of state $|\Psi_{final}\rangle$ is denoted by red solid curve, and probability of intermediate states are denoted by red grey curve. Panel (b) shows the evolution without consideration of electron spin transition. In (b), probability of state $|\Psi_{initial}'\rangle$ is denoted by cyan solid curve. Other curves represent as same in panel (a). Inserted figures in both panels show the curves of time-dependent probabilities of subspaces $\mathcal{A}$ and $\mathcal{D}$. Probability of $\mathcal{A}$ is denoted by red solid curve, and probability of $\mathcal{D}$ is denoted by blue solid curve.}
		\label{fig:WithoutWithSpinFlip}
	\end{figure*}
	
	The coupling strength of photon and the electron in the cavity takes the form:
	\begin{equation}
		\label{eq:Interaction}
		g_n=\sqrt{\hbar\omega_n/V}dE\left(x\right)
	\end{equation}
	where $\omega_n$ is transition frequency, $V$ is the effective volume of the cavity, $d$ is the dipole moment of the transition between the ground and the perturbed states and $E\left(x\right)$ describes the spatial arrangement of the atom in the cavity, which has the form $E\left(x\right) = sin\left(\pi x/l\right)$, here $l$ is the length of the cavity. To ensure the confinement of the photon in the cavity, $l$ has to be chosen such that $l = r\lambda/2$ is a multiple of the photon wavelength $\lambda$. In experiments, $r=1$ is often chosen to decrease the effective volume of the cavity, which makes it possible to obtain dozens of Rabi oscillations \cite{Rempe1987}. We assume that $\Omega^c<\Omega^s<\omega^{\uparrow}=\omega^{\downarrow}<\Omega^{\uparrow}=\Omega^{\downarrow}$, thus $g_{\Omega^c}<g_{\Omega^s}<g_{\omega^{\uparrow}}=g_{\omega^{\downarrow}}<g_{\Omega^{\uparrow}}=g_{\Omega^{\downarrow}}$ according to Eq. \eqref{eq:Interaction}.
	
	In simulations: 
	
	$\Omega^{\uparrow}=\Omega^{\downarrow}$, $\omega^{\uparrow}=\omega^{\downarrow}=0.5*\Omega^{\uparrow}$, $\Omega^s=0.1*\Omega^{\uparrow}$, $\Omega^c=0.01*\Omega^{\uparrow}$;
	
	$g_{\Omega^{\uparrow}}=g_{\Omega^{\downarrow}}=0.01*\Omega^{\uparrow}$, $g_{\omega^{\uparrow}}=g_{\omega^{\downarrow}}=0.5*g_{\Omega^{\uparrow}}$, $g_{\Omega^s}=0.1*g_{\Omega^{\uparrow}}$, $g_{\Omega^c}=0.05*g_{\Omega^{\uparrow}}$;
	
	$\zeta=0.5*g_{\Omega^{\uparrow}}$, $\zeta_2=10*g_{\Omega^{\uparrow}}$, $\zeta_1=g_{\Omega^{\uparrow}}$, $\zeta_0=0$.
	
	In Markovian open systems, we assume that the dissipative rates of all types of photon leakage are equal:
	
	$\gamma_{\omega^{\uparrow}}=\gamma_{\omega^{\downarrow}}=\gamma_{\Omega^{\uparrow}}=\gamma_{\Omega^{\downarrow}}=\gamma_{\Omega^s}=\gamma_{\Omega^c}=0.1*g_{\Omega^{\uparrow}}$.
	
	\subsection{Without consideration of electron spin transition} 
	\label{subsec:WithoutSpinFlip}
	
	In this subsection, a photon with mode $\Omega^{\uparrow}$ and a photon with mode $\Omega^{\downarrow}$ are the only ones pumped into the system at the beginning, corresponding to the initial state $|\Psi_{initial}\rangle$, described in Fig. \ref{fig:SpinTrans}(a), where two electrons with $\downarrow$ are in atomic ground state of different atoms, and photon with mode $\Omega^s$, which can excite electron from $\downarrow$ to $\uparrow$, is absent. Electron spin transition is thus prohibited. Additionally, only the influx of photons with modes $\Omega^{\uparrow}$ and $\Omega^{\downarrow}$ is taken into account. The influx of photons with modes $\omega^{\uparrow}$ and $\omega^{\downarrow}$ is forbidden. And as stated in Sec. \ref{sec:Thermally}, the influx rate is always lower than the corresponding dissipative rate.
	
	We assume that $\mu_{\omega^{\uparrow}}=\mu_{\omega^{\downarrow}}=0,\ \mu_{\Omega^{\uparrow}}=\mu_{\Omega^{\downarrow}}=0.5$.
	
	In Fig. \ref{fig:SpinTrans} two electrons with different spins are anchored in the molecular ground orbital, describing the $|\Psi_{final}\rangle$. Theoretically, it is impossible to accomplish $|\Psi_{final}\rangle$ because hybridization of atomic orbitals only occurs when two electrons in identically excited atomic orbitals have different spins. The red solid curve representing $|\Psi_{final}\rangle$ is always equal to 0 during the whole evolution, as shown by the numerical results in Fig. \ref{fig:WithoutWithSpinFlip}(a). And in inserted figure, red solid curve representing $\mathcal{A}$, which is the sum of probabilities of all states belonging to associative system corresponding to $|0\rangle_n$, is also always equal to 0. And blue solid curve representing $\mathcal{D}$, which is the sum of probabilities of all states belonging to dissociative system corresponding to $|1\rangle_n$, is always equal to 1. This indicates that no energy enters the associative system throughout evolution in our model, and the entire system remains completely dissociated. This means that when two electrons are both fixed with the same spin, and when electron spin transition is inhibitive, formation of the neutral hydrogen molecule is impossible.
	
	\begin{figure*}
		\centering
		\includegraphics[width=1.\textwidth]{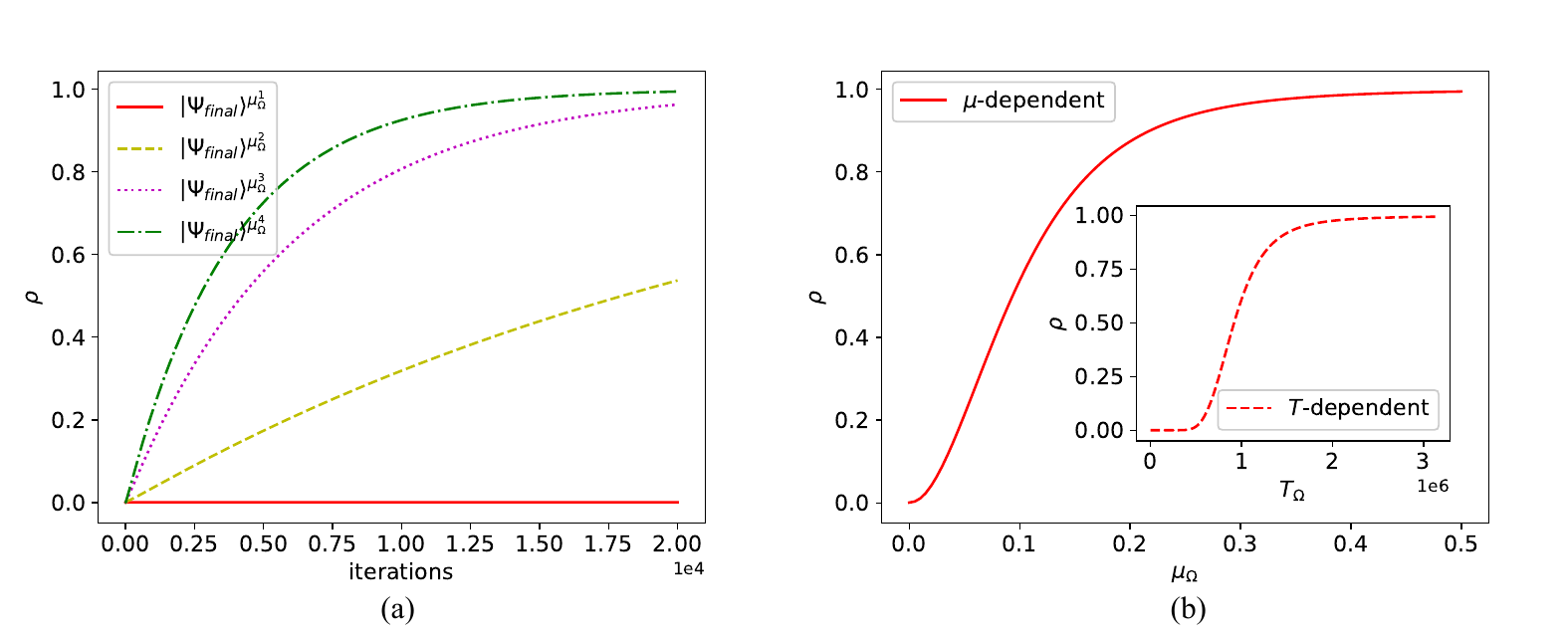}
		\caption{(online color) {\it Temperature variation of photonic modes $\Omega^{\uparrow}$ and $\Omega^{\downarrow}$.} In panel (a), time-dependent curves of $|\Psi_{final}\rangle$ are corresponding to $\mu_{\Omega^1}$ (red solid), $\mu_{\Omega^2}$ (yellow dashed), $\mu_{\Omega^3}$ (magenta dotted) and $\mu_{\Omega^4}$ (green dash--dotted), respectively. In panel (b), red solid curve represents the probability of $|\Psi_{final}\rangle$ when iterations reaches $20000$, with the increase of $\mu_{\Omega}$ from $0$ to $0.5$. Red dashed curve in inserted figure represents the $T$-dependent probability of $|\Psi_{final}\rangle$ when iterations reaches $20000$.}
		\label{fig:TemperatureAtom}
	\end{figure*}
	
	\subsection{With consideration of electron spin transition} 
	\label{subsec:WithSpinFlip}
	
	The association--dissociation model of the neutral hydrogen molecule now includes the spin-flip photon, and electron spin transition is a possibility. According to Fig. \ref{fig:SpinTrans}(b), the initial state is $|\Psi_{initial}'\rangle$, where a photon with the mode $\Omega^s$ is introduced. We further stipulate that an electron can only undergo an electron spin transition when it is in an atomic state, both excited and ground.
	
	Similarly, for added photon with mode $\Omega^s$, we assume that $\mu_{\Omega^s}=0.5$. Others as same in Subsec. \ref{subsec:WithoutSpinFlip}.
	
	According to numerical results in Fig. \ref{fig:WithoutWithSpinFlip}(b), we discovered that the red solid curve $|\Psi_{final}\rangle$ climbs and reaches $1$ at the end when electron spin transition is taken into account. It indicates that the formation of H$_2$ has been accomplished and that there are no longer any free hydrogen atoms. Additionally, the red solid curve $\mathcal{A}$ rises and reaches $1$ in inserted figure, while the blue solid curve $\mathcal{D}$ declines to $0$. In other words, when electron spin transition is allowed, the formation of a neutral hydrogen molecule is conceivable when two electrons have different spins.
	
	We make the assumption that the dissipative rate of all types of photons is the same, which is why what is depicted in Fig. \ref{fig:WithoutWithSpinFlip}(b) is accurate. $\mu_{\Omega^{\uparrow}},\ \mu_{\Omega^{\downarrow}}$ and $\mu_{\Omega^s}$ are equal to $0.5$, and $\mu_{\omega^{\uparrow}},\ \mu_{\omega^{\downarrow}}$ --- $0$ (which means that the inflow rates of photons with the modes $\omega^{\uparrow}$ and $\omega^{\downarrow}$ are both $0$). In order to force electrons to move from the molecular ground orbital to the excited orbital, as shown in Fig. \ref{fig:AssDissModel}(d), the decomposition of hydrogen molecules must absorb photons with modes $\omega^{\uparrow}$ and $\omega^{\downarrow}$, but because these photons cannot be replenished, they will gradually leak until they are completely absent in the cavity. As a result, the system finally evolves over time to generate a stable neutral hydrogen molecule.
		
	\subsection{Temperature variation} 
	\label{subsec:Temperature}
	
	We are currently looking at how changes in temperature affect the evolution and the formation of neutral hydrogen molecules using the photonic modes $\Omega^{\uparrow}$, $\Omega^{\downarrow}$, $\omega^{\uparrow}$, $\omega^{\downarrow}$ and $\Omega^s$.
	
	In this subsection we use $\mu$ instead of temperature $T$ as the abscissa, and for convenience suppose $\mu_{\Omega}=\mu_{\Omega^{\uparrow}}=\mu_{\Omega^{\downarrow}}$ and $\mu_{\omega}=\mu_{\omega^{\uparrow}}=\mu_{\omega^{\downarrow}}$.
	
	\subsubsection*{Temperature variation of photonic modes $\Omega^{\uparrow}$ and $\Omega^{\downarrow}$}
	\label{subsubsec:TemVarOmega}
	
	\begin{figure*}
		\centering
		\includegraphics[width=1.\textwidth]{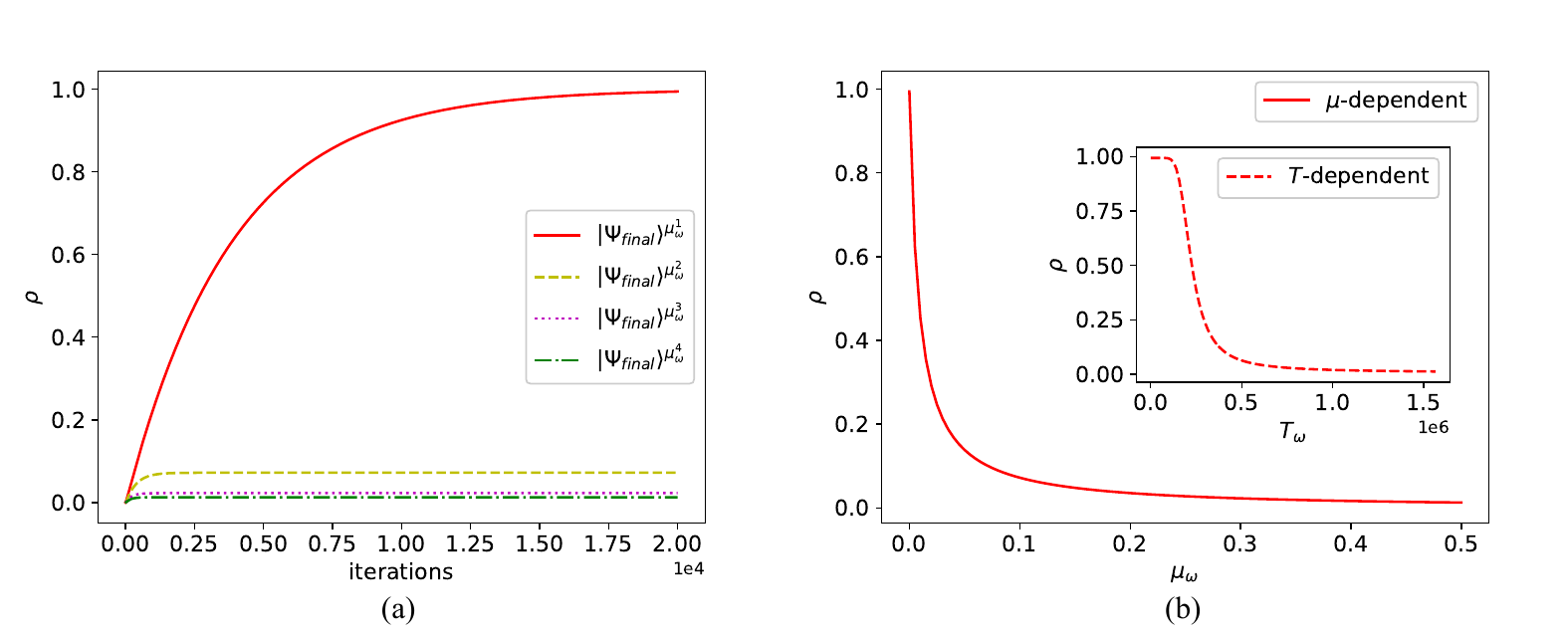}
		\caption{(online color) {\it Temperature variation of photonic modes $\omega^{\uparrow}$ and $\omega^{\downarrow}$.} In panel (a), time-dependent curves of $|\Psi_{final}\rangle$ are corresponding to $\mu_{\omega^1}$ (red solid), $\mu_{\omega^2}$ (yellow dashed), $\mu_{\omega^3}$ (magenta dotted) and $\mu_{\omega^4}$ (green dash--dotted), respectively. In panel (b), red solid curve represents the probability of $|\Psi_{final}\rangle$ when iterations reaches $20000$, with the increase of $\mu_{\omega}$ from $0$ to $0.5$. Red dashed curve in inserted figure represents the $T$-dependent probability of $|\Psi_{final}\rangle$ when iterations reaches $20000$.}
		\label{fig:TemperatureMolecule}
	\end{figure*}
	
	\begin{figure}
		\centering
		\includegraphics[width=0.5\textwidth]{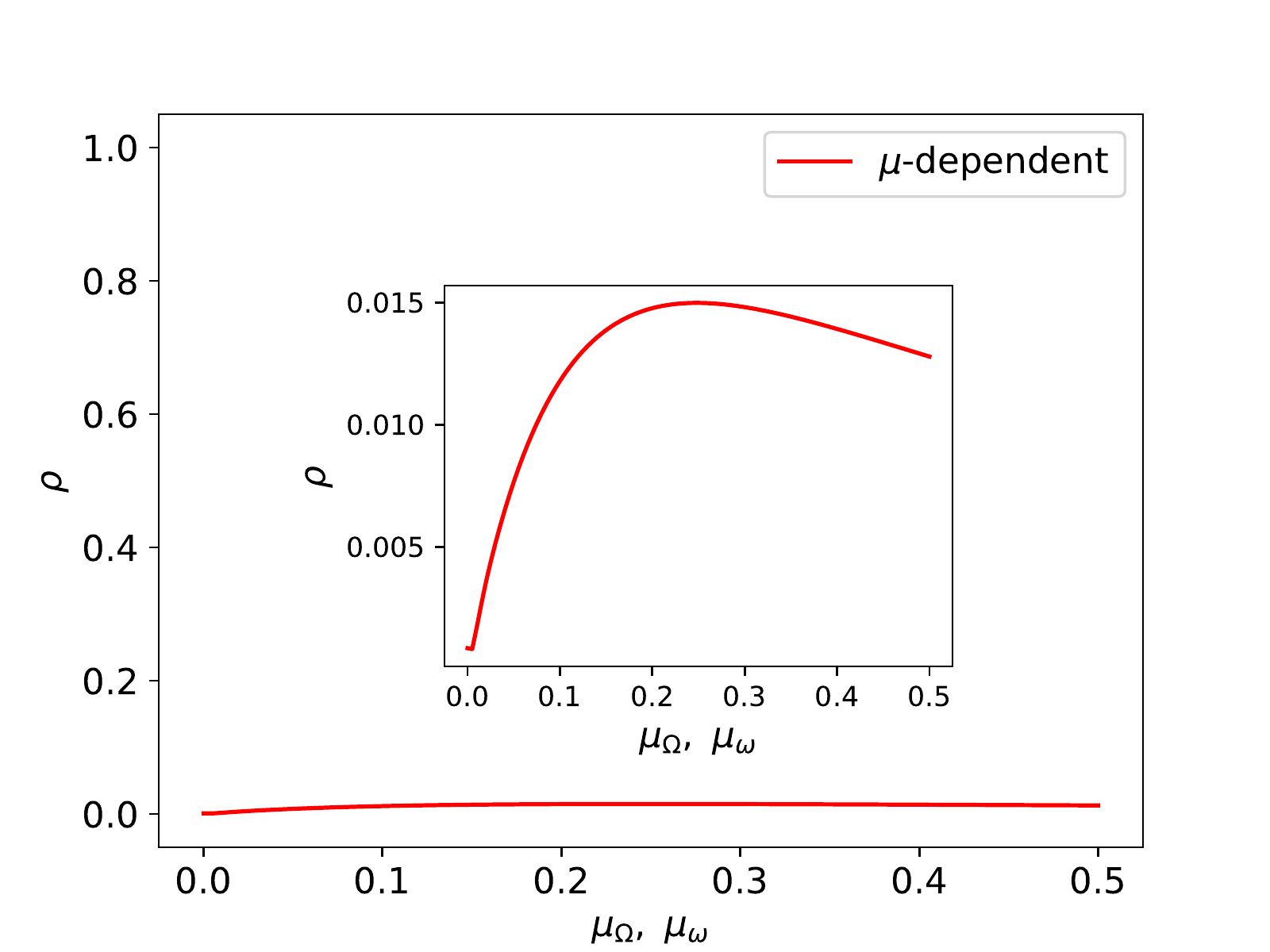}
		\caption{(online color) {\it Counteraction of temperature variations of atomic and molecular photonic modes.} Red curve represents the probability of $|\Psi_{final}\rangle$ when iterations reaches $20000$, with the simultaneous increases of $\mu_{\Omega}$ and $\mu_{\omega}$ from $0$ to $0.5$.}
		\label{fig:TemperatureAtomMolecule}
	\end{figure}
	
	We assume that $\mu_{\omega}=0,\ \mu_{\Omega^s}=0.5$.
	
	In Fig. \ref{fig:TemperatureAtom}(a), we chose four instances that vary in various $\mu_{\Omega}$: $\mu_{\Omega}^1=0,\ \mu_{\Omega}^2=0.1,\ \mu_{\Omega}^3=0.3,\ \mu_{\Omega}^4=0.5$. We discovered that neutral hydrogen molecule forms more quickly the higher the $\mu_{\Omega}$ (or $T_{\Omega}$). The circumstance where $\mu_{\Omega}^1=0$ (in this case, $T_{\Omega}^1=0K$) occurs is where formation moves the slowest, indicated by red solid curve. The fastest formation occurs when $\mu_{\Omega}^4=0.5$, indicated by green dash--dotted curve. The probability of the $|\Psi_{final}\rangle$ never approaches $1$ when the $\mu_{\Omega}$ is equal to $0$. However, once $\mu_{\Omega}$ is bigger than $0$, the probability of $|\Psi_{final}\rangle$ will reach $1$ as long as the duration is long enough. Because molecule photons are not renewed, atomic photons are continually being added back into the system. Therefore, the entire system will progressively change in order to produce a stable molecular state.
	
	We now raise $\mu_{\Omega}$ from $0$ to $0.5$. In each case we take the value of final state when the number of iterations reaches $20000$. We can intuitively perceive the trend of $|\Psi_{final}\rangle$ with the growth of $\mu_{\Omega}$ in Fig. \ref{fig:TemperatureAtom}(b). Probability of $|\Psi_{final}\rangle$ is close to $0$ when $\mu_{\Omega}$ is near to $0$. It begins to expand slowly as the $\mu_{\Omega}$ rises, then quickly accelerates until it reaches a top, which is close to $1$. From the inserted figure in Fig. \ref{fig:TemperatureAtom}(b), we can see that the $T$-dependent curve of probability has the same trend as the $\mu$-dependent curve, but there is a hysteresis near $0K$.
	
	\subsubsection*{Temperature variation of photonic modes $\omega^{\uparrow}$ and $\omega^{\downarrow}$}
	\label{subsubsec:TemVaromega}
	
	We assume that $\mu_{\Omega}=\mu_{\Omega^s}=0.5$.
	
	In Fig. \ref{fig:TemperatureMolecule}(a), we chose four instances that vary in various $\mu_{\omega}$: $\mu_{\omega}^1=0,\ \mu_{\omega}^2=0.1,\ \mu_{\omega}^3=0.3,\ \mu_{\omega}^4=0.5$. And in Fig. \ref{fig:TemperatureMolecule}(b) we increase $\mu_{\omega}$ rises from $0$ to $0.5$.
	
	It is clear from Fig. \ref{fig:TemperatureMolecule} that the temperature variation of molecular photonic modes affects neutral evolution and hydrogen molecule formation in the opposite way from atomic photonic modes: the higher $\mu_{\omega}$ (or $T_{\omega}$), the slower evolution and formation.
	
	The probability of the $|\Psi_{final}\rangle$ can reach $1$ only when the $\mu_{\omega}$ is $0$, which is different from the Fig. \ref{fig:TemperatureAtom}(a). Even if the duration is long enough, when the $\mu_{\omega}$ is not $0$, the probability cannot increase to $1$. The system will reach equilibrium between associative and dissociative systems because $\mu_{\Omega}$ and $\mu_{\omega}$ are both non-zero numbers at this point, meaning that the atomic and molecular photons are replenished simultaneously (although the replenishing efficiencies may differ). The value of the $|\Psi_{final}\rangle$ probability at equilibrium depends on the ratio of $\mu_{\Omega}$ and $\mu_{\omega}$. For molecular photon, the $T$-dependent curve of probability has a hysteresis, too.
	
	\subsubsection*{Counteraction of temperature variations of atomic and molecular photonic modes.}
	\label{subsubsec:Counteraction}
	
	\begin{figure*}
		\centering
		\includegraphics[width=1\textwidth]{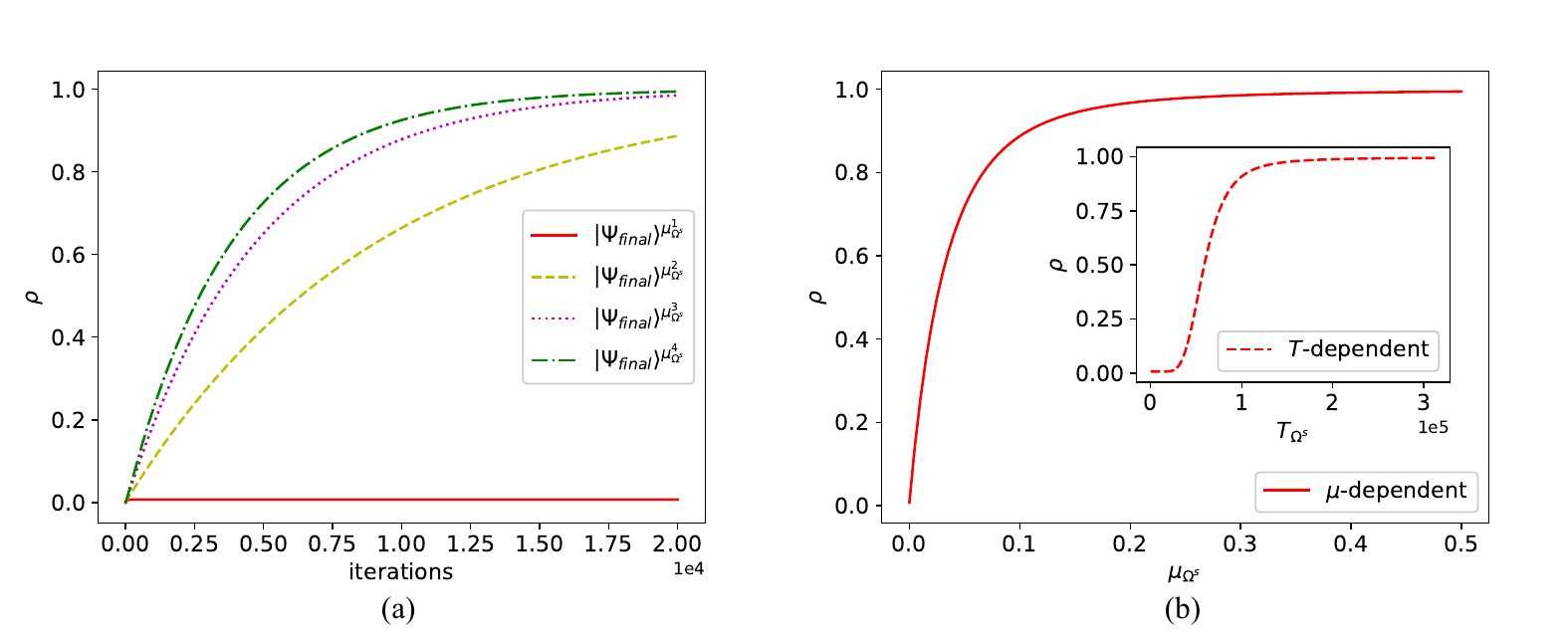}
		\caption{(online color) {\it Temperature variation of photonic mode $\Omega^s$.} In (a), time-dependent curves of $|\Psi_{final}\rangle$ are corresponding to $\mu_{\Omega^s}^1$ (red solid), $\mu_{\Omega^s}^2$ (yellow dashed), $\mu_{\Omega^s}^3$ (magenta dotted) and $\mu_{\Omega^s}^4$ (green dash--dotted), respectively. In (b), red solid curve represents the probability of $|\Psi_{final}\rangle$ when iterations reaches $20000$, with the increases of $\mu_{\Omega^s}$ from $0$ to $0.5$. Red dashed curve in inserted figure represents the $T$-dependent probability of $|\Psi_{final}\rangle$ when iterations reaches $20000$.}
		\label{fig:TemperatureElectronSpin}
	\end{figure*}
	
	We said that temperature variation of atomic photonic modes is positive effect to evolution and formation of neutral hydrogen molecule, and temperature variation of molecular photonic modes is negative.
	
	Now we consider these both opposite effects at the same time. We assume that $\mu_{\Omega^s}=0.5$. And we increase $\mu_{\Omega},\ \mu_{\omega}$ both rises from $0$ to $0.5$. Specially, $\mu_{\Omega}$ are always equal to $\mu_{\omega}$.
	
	The probability of the $|\Psi_{final}\rangle$ curve in Fig. \ref{fig:TemperatureAtomMolecule} is practically equal to zero as the atomic and molecular temperatures rise. Despite the curve's apparent small oscillation between the intervals $\left[0,\ 0.016\right]$ in the inset graphic, we choose to ignore it. The initial state $|\Psi_{initial}'\rangle$, shown in Fig. \ref{fig:SpinTrans}(b), is not an equilibrium state between associative and dissociative systems, which accounts for the mild oscillations.
	
	Thus, it is impossible for a neutral hydrogen molecule to form since the effects of temperature change on atomic and molecular photons cancel each other out.
	
	\subsubsection*{Temperature variation of photonic modes $\Omega^s$}
	\label{subsubsec:TemVarOmegaS}
	
	We assume that $\mu_{\omega}=0,\ \mu_{\Omega}=0.5$.
	
	In Fig. \ref{fig:TemperatureElectronSpin}(a), we also chose four instances that vary in various $\mu_{\Omega^s}$: $\mu_{\Omega^s}^1=0,\ \mu_{\Omega^s}^2=0.1,\ \mu_{\Omega^s}^3=0.3,\ \mu_{\Omega^s}^4=0.5$. We found that the higher $\mu_{\Omega^s}$ (or $T_{\Omega^s}$), the faster formation of neutral hydrogen molecule. When $\mu_{\Omega^s}^1=0$ (here $T_{\Omega^s}^1=0K$), denoted by red solid curves, formation is slowest among all situations. When $\mu_{\Omega^s}^4=0.5$, denoted by green dash--dotted curves, formation is fastest among all situations. Same as $\mu_{\Omega}$, when the $\mu_{\Omega^s}$ is equal to $0$, the probability of $|\Psi_{final}\rangle$ never reaches $1$. But once $\mu_{\Omega^s}$ is greater than 0, then as long as the time is long enough, probability of $|\Psi_{final}\rangle$ will reach $1$.
	
	Now we increase $\mu_{\Omega^s}$ from $0$ to $0.5$. In each case we take the value of final state when the number of iterations reaches $20000$. In Fig. \ref{fig:TemperatureElectronSpin}(b), when $\mu_{\Omega^s}$ rises, probability of $|\Psi_{final}\rangle$ increases immediately abruptly. When $\mu_{\Omega^s}$ is larger enough, probability of $|\Psi_{final}\rangle$ reaches top, which is close to $1$.For spin-flip photon, the $T$-dependent curve of probability also has a hysteresis near $0K$ like those in Fig. \ref{fig:TemperatureAtom}(b).
	
	\subsection{With covalent bond and phonon} 
	\label{subsec:BondPhonon}
	
	\begin{figure}
		\centering
		\includegraphics[width=0.5\textwidth]{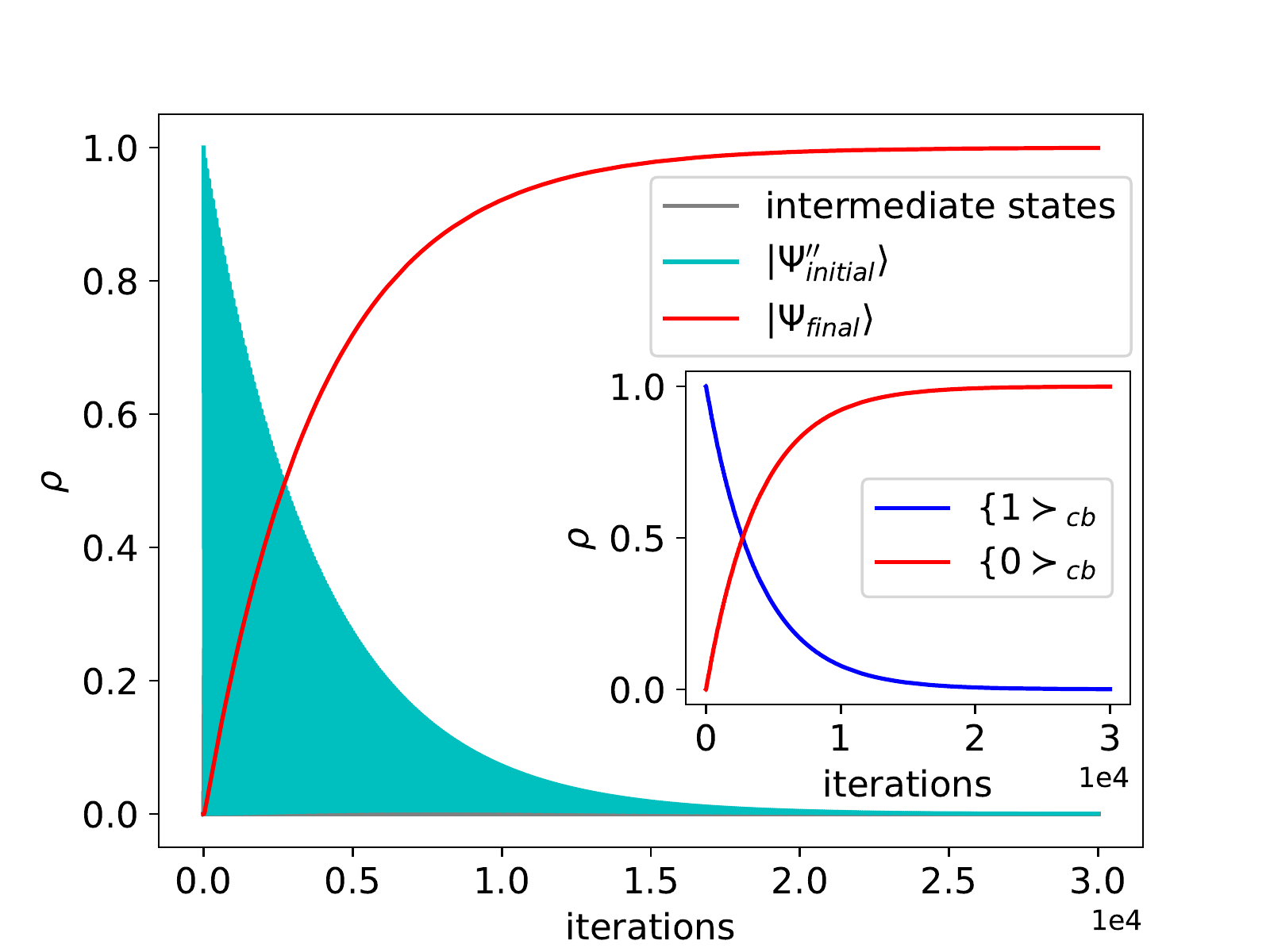}
		\caption{(online color) {\it The evolution with consideration of covalent bond and phonon.} Probability of state $|\Psi_{initial}''\rangle$ is denoted by cyan solid curve and probability of state $|\Psi_{final}\rangle$ is denoted by red solid curve. Inserted figure shows the curves of time-dependent probabilities of $\left\{0\succ_{cb}\right.$ and $\left\{1\succ_{cb}\right.$. Probability of $\left\{0\succ_{cb}\right.$ is denoted by red solid curve, and probability of $\left\{1\succ_{cb}\right.$ is denoted by blue solid curve.}
		\label{fig:WithBondPhonon}
	\end{figure}
	
	Now we introduce covalent bond and phonon in the association--dissociation model of neutral hydrogen molecule. Initial state is $|\Psi_{initial}''\rangle$, described in Fig. \ref{fig:BondPhonon}.
	
	According to numerical results in Fig. \ref{fig:WithBondPhonon}, we found red solid curve $|\Psi_{final}\rangle$ rises and reaches $1$ at the end. And in inserted figure, red solid curve $\left\{0\succ_{cb}\right.$ (as same as $\mathcal{A}$) also rises and reaches 1, and blue solid curve $\left\{1\succ_{cb}\right.$ (as same as $\mathcal{D}$) descends to 0. These results are consistent with Fig. \ref{fig:WithoutWithSpinFlip}(b). But this model is more straightforward and understandable.
	
	\section{Concluding discussion and future work} 
	\label{sec:ConcluFuture}
	
	In this paper, we simulate the association of the neutral hydrogen molecule in the cavity QED model --- the TCHM. The association--dissociation model has been constructed, and several analytical findings have been drawn from it:
	
	In Secs. \ref{subsec:WithoutSpinFlip} and \ref{subsec:WithSpinFlip}, we proved hybridization of atomic orbitals and formation of neutral hydrogen molecule only happens when electrons with different spins. Then the influence of variation of $T_{\Omega^{\uparrow}}$, $T_{\Omega^{\downarrow}}$, $T_{\omega^{\uparrow}}$, $T_{\omega^{\downarrow}}$ and $T_{\Omega^s}$ to the evolution and the formation of neutral hydrogen molecule is obtained in Sec. \ref{subsec:Temperature}: for $T_{\Omega^{\uparrow}}\left(T_{\Omega^{\downarrow}}\right)$ and $T_{\Omega^s}$, the higher temperature, the faster neutral hydrogen molecule formation; for $T_{\omega^{\uparrow}}\left(T_{\omega^{\downarrow}}\right)$, the higher temperature, the slower neutral hydrogen molecule formation. Finally, we studied the more accurate model with covalent bond and phonon in Sec. \ref{subsec:BondPhonon}.
	
	Although our approach is still imperfect, it has the advantages of being simple and scalable. It will be more subdued in this manner. Additionally, this model can be modified in the future for use with more intricate chemical and biologic models.
	
	\begin{acknowledgments}
	The reported study was funded by China Scholarship Council, project number 202108090483. The authors acknowledge Center for Collective Usage of Ultra HPC Resources (https://www.parallel.ru/) at Lomonosov Moscow State University for providing supercomputer resources that have contributed to the research results reported within this paper.
	\end{acknowledgments}

\bibliography{bibliography}

\onecolumngrid

\appendix

	\section{Complete expressions for TCHM}
	\label{appx:ComExpTCH}
	
	\subsection{TCM}
	\label{appx:TC}
	
	We consider the TCM to describe the interaction of atomic ensembles ($N$ atoms) with photons in an optical cavity (the simplest model with a two-level atom, called JCM, is shown in Fig. \ref{appxfig:TCandTCH}(a)). Hamiltonian of TCM for the weak interaction $g\ll\hbar\omega_c\approx\hbar\omega_a$ (RWA) looks as follows
	\begin{subequations}
		\label{appxeq:TCHamil}
		\begin{align}
			H_{TC}&=\underbrace{\hbar\omega_c a^{\dag}a}_{H_{field}}+\underbrace{\hbar\omega_a\sum_{i=1}^N\sigma_i^{\dag}\sigma_i}_{H_{atoms}}+\underbrace{\sum_{i=1}^Ng_i\left(a^{\dag}+a\right)\left(\sigma_i^{\dag} + \sigma_i\right)}_{H_{int}}\label{appxeq:TC}\\
			H_{TC}^{RWA}&=\hbar\omega_c a^{\dag}a+\hbar\omega_a\sum_{i=1}^N\sigma_i^{\dag}\sigma_i+\sum_{i=1}^Ng_i\left(a^{\dag}\sigma_i+a\sigma_i^{\dag}\right)\label{appxeq:TCRWA}
		\end{align}
	\end{subequations}
	
	\subsection{TCHM}
	\label{appx:TCH}
	
\begin{figure}
		\centering
		\includegraphics[width=1\textwidth]{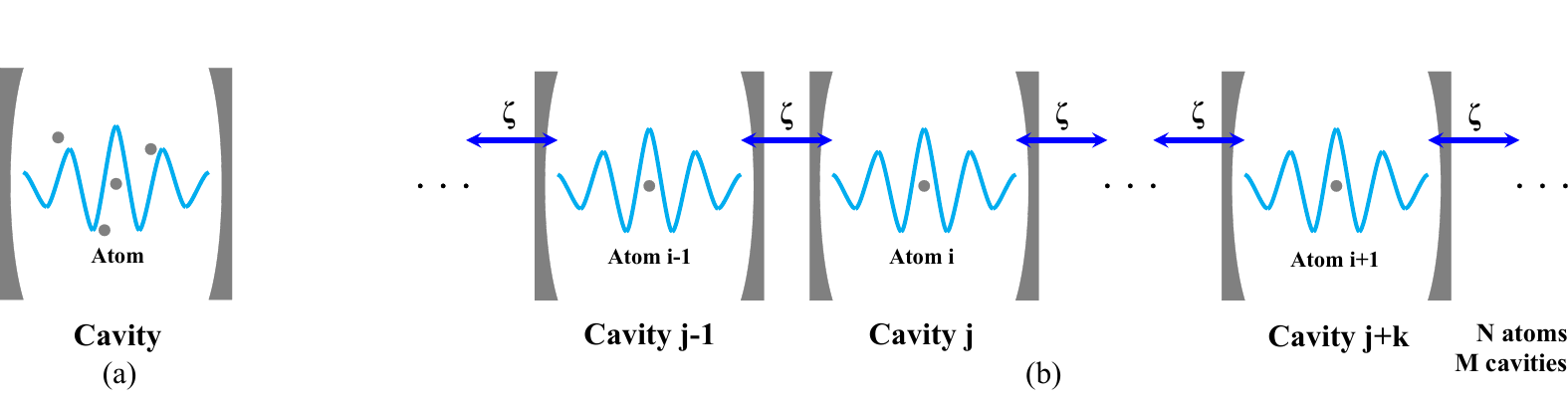}
		\caption{(online color) {\it TCM and TCHM.} TCM with N two-level atoms in an optical cavity is shown in panel (a), TCHM with N two-level atoms and M optical cavities coupled by an optical fibre is shown in panel (b). Atoms are denoted by grey dots.}
		\label{appxfig:TCandTCH}
	\end{figure}

	TCM has been generalized to several cavities coupled by an optical fibre --- TCHM in Fig. \ref{appxfig:TCandTCH}(b). Photons can move between optical cavities through optical fibres. Hamiltonian of TCHM for RWA looks as follows
	\begin{subequations}
		\label{appxeq:TCHHamil}
		\begin{align}
			H_{TCH}&=\sum_{j=1}^M\left\{\underbrace{\hbar\omega_{c_j}a_j^{\dag}a_j+\hbar\omega_{a_j}\sum_{i=1}^N\sigma_{i_j}^{\dag}\sigma_{i_j}+\sum_{i=1}^Ng_i\left(a_j^{\dag}+a_j\right)\left(\sigma_{i_j}^{\dag} + \sigma_{i_j}\right)}_{H_{TC}}\right\}+\zeta\sum_{j=1}^M\left(a_{j+1}^{\dag}a_j+a_j^{\dag}a_{j+1}\right)\label{appxeq:TCH}\\
			H_{TCH}^{RWA}&=\sum_{j=1}^M\left\{\hbar\omega_{c_j}a_j^{\dag}a_j+\hbar\omega_{a_j}\sum_{i=1}^N\sigma_{i_j}^{\dag}\sigma_{i_j}+\sum_{i=1}^Ng_i\left(a_j^{\dag}\sigma_{i_j}+a_j\sigma_{i_j}^{\dag}\right)\right\}+\zeta\sum_{j=1}^M\left(a_{j+1}^{\dag}a_j+a_j^{\dag}a_{j+1}\right)\label{appxeq:TCHRWA}
		\end{align}
	\end{subequations}
	where $M$ --- number of optical cavities, $\zeta$ --- atoms leap strength (tunnelling strength) between neighbouring cavities.
	
	\section{Theorem for thermally stationary state}
	\label{appx:Theorem}
	
	{\bf Theorem} Thermally stationary state of atoms and field at the temperature T has the form
	\begin{equation}
		\label{appxeq:Space}
		\rho_{stat}=\rho_{ph}\otimes\rho_{at}
	\end{equation}		
where $\rho_{at}$ is the state of atoms and the state of field $\rho_{ph}={\cal G}\left(T\right)_f$ is equilibrium state at this temperature.

	{\bf Proof} We expand Hamiltonian $H=H_{at}+H_{ph}$ to the atomic part $H_{at}$ and purely photonic component $H_{ph}=\hbar\omega a^{\dag}a$, and introduce notations $U_{dt}\left(\rho\right)=e^{-\frac{i}{\hbar}H_{at}dt}\rho e^{\frac{i}{\hbar}H_{at}dt}$, $U_{dt}'\left(\rho\right)=e^{-\frac{i}{\hbar}H_{ph}dt}\rho e^{\frac{i}{\hbar}H_{ph}dt}$ for the action of summands of the unitary part of the Lindblad superoperator $\mathcal{L}\left(\rho\right)$ in Eq. \eqref{eq:QME} to the density matrix on the short time segment $dt$.
	
	We denote through $L_{dt}'\left(\rho\right)=\rho+idtL\left(\rho\right)$ the action of Lindblad superoperator on the density matrix in the time $dt$. With accuracy $dt$ we then have the approximate equation
	\begin{equation}
		\label{appxeq:Approx}
		\rho\left(t\right)\approx\left(U_{dt}U_{dt}'L_{dt}'\right)^{\frac{t}{dt}}\left(\rho\right)
	\end{equation}
	analogous to the Trotter formula, which comes from Euler method of the solution of quantum master equation in Eq. \eqref{eq:QME}.
	
	Since operators $L_{dt}',\ U_{dt}'$ act on the photon component of state only, and $U_{dt}$ --- on the photon and atomic components, the stationary state $\rho_{stat}$ at the randomly chosen constants of interaction $g,\ \gamma_{k'},\ \gamma_k$ must not change after the action of operators a $L_{dt}''=U_{dt}'L_{dt}'$, and operator $Udt$.
	
	We fix arbitrary basic state of atoms $I,\ J$ and consider the minor $\rho_{IJ}$ of the matrix $\rho$, formed by coefficients at the basic states $|I,i\rangle\langle J,j|$, where $|i\rangle,\ |j\rangle$ --- are Fock state of the field. The operator $L_{dt}'$, acting on the photon states, factually acts on the minor $\rho_{IJ}$.
	
	We will denote by the sign $\tilde{\rho}$ the result of the application of operator $L_{dt}'$ to the minor $\rho_{IJ}=\rho$, so that $\tilde{\rho}_{ij}$ denote mtrix elements of this result and $\rho_{ij}$ --- matrix elements of the initial state $\rho$; we will enumerate rows and columns of this matrix beginning with zero, so that $i,j=0,1,2,\cdots$, and omit in the notation atomic states $I,J$, which will always be the same. Taking into account the definition of operators of the creation and annihilation of photons, we have
	\begin{equation}
		\label{appxeq:Rhoij}
		\tilde{\rho}_{ij}=\rho_{ij}+{\gamma_k}\left(\sqrt{i+1}\sqrt{j+1}\rho_{i+1,j+1}-\frac{i+j}{2}\rho_{ij}\right)+{\gamma_{k'}}\left\{\sqrt{i}\sqrt{j}\rho_{i-1,j-1}-\left(\frac{i+j}{2}+1\right)\rho_{ij}\right\}
	\end{equation}
	We also have $L_{dt}''\left(\rho\right)=\rho,\ L_{dt}''\left(\tilde{\rho}\right)=\tilde{\rho}$. The operator $U_{dt}'$ does not change the diagonal members of the matrix, and it multiplies nondiagonal members to the coefficient $e^{\pm i\omega\left(i-j\right)dt}$. Because the coefficient $\omega$, determining the phase is not connected with $\gamma_{k'}$ and $\gamma_k$ from the Eq. \eqref{appxeq:Rhoij}, this multiplication cannot compensate in the first order on $dt$ the addition to $\rho_{ij}$ from Eq. \eqref{appxeq:Rhoij}, and hence in the matrix $\rho$ nondiagonal members are zero. We consider the diagonal of this matrix. From the Eq. \eqref{appxeq:Rhoij} we have
	\begin{equation}
		\label{appxeq:Rhoii}
		\tilde{\rho}_{ii}=\rho_{ii}+{\gamma_k}\left\{\left(i+1\right)\rho_{i+1,i+1}-i\rho_{ii}\right\}+{\gamma_{k'}}\left\{i\rho_{i-1,i-1}-\left(i+1\right)\rho_{ii}\right\}
	\end{equation}
	From the Eq. \eqref{appxeq:Rhoii} we can obtain the recurrent equation for the elements of diagonal, but it is possible to get their form easier. We apply to the diagonal of $\rho$ the representation of quantum hydrodynamics. The flow of probability from the basic state $|i\rangle\langle i|$ to the state $|i+1\rangle\langle i+1|$ is $\left(i+1\right)\rho_{ii}\gamma_{k'}$, and the reverse flow is $\left(i+1\right)\rho_{i+1,i+1}\gamma_k$, from which we get that $\rho_{ii}$ is proportional to $\mu^i$, that is required.

	Now we substitute this expression for the diagonal element to the Eq. \eqref{appxeq:Rhoii}, and obtain $\tilde{\rho_{ii}}=\rho_{ii}$. Since the choice of $I,\ J$ was arbitrary, we obtain $\rho_{stat}={\cal G}\left(T\right)_f\otimes\rho_{at}$, that is required. Theorem is proved.
	
	\section{Tensor product and generator algorithm}
	\label{appx:TensorGenerator}

\begin{figure}
		\centering
		\includegraphics[width=1\textwidth]{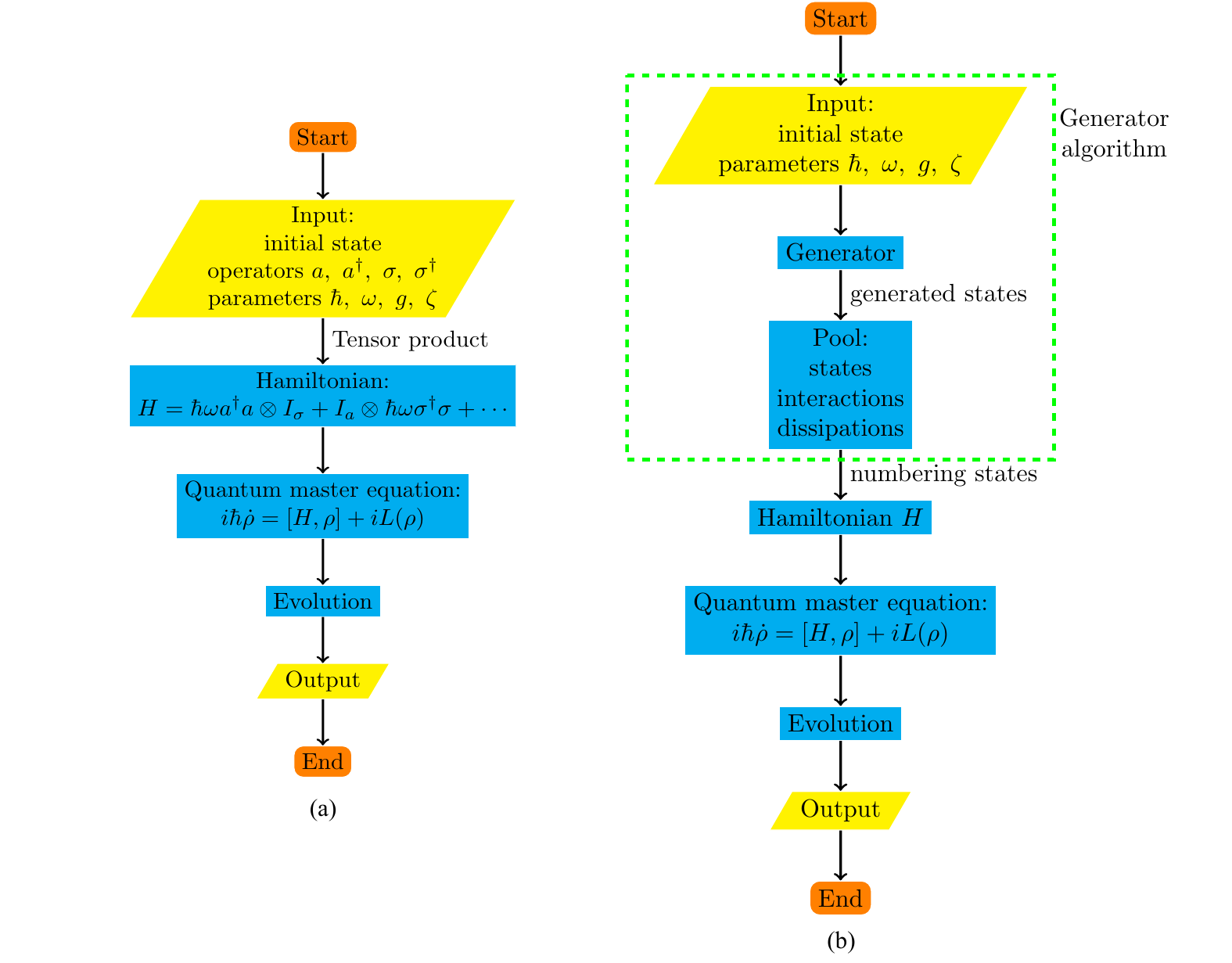}
		\caption{(online color) {\it Tensor product and generator algorithm.}}
		\label{appxfig:TensorGenerator}
	\end{figure}

	Now we consider the simplest TCM with only one two-level atom, which is shown in Fig. \ref{appxfig:TCandTCH}(a). And its Hamiltonian with RWA takes following form
	\begin{equation}
		\label{appxeq:TensorProduct}
		\begin{aligned}
			H_{TC}^{RWA}&=\hbar\omega a^{\dag}a+\hbar\omega\sigma^{\dag}\sigma+g\left(a^{\dag}\sigma+a\sigma^{\dag}\right)\\
			&=\hbar\omega a^{\dag}a\otimes I_{\sigma}+I_a\otimes\hbar\omega\sigma^{\dag}\sigma+g\left\{\left(a^{\dag}\otimes I_{\sigma}\right)\left(I_a\otimes\sigma\right)+\left(a\otimes I_{\sigma}\right)\left(I_a\otimes\sigma^{\dag}\right)\right\}\\
			&=\hbar\omega\begin{array}{c@{\hspace{-5pt}}l}
			 \begin{array}{c}
			 	|0\rangle \\
			 	|1\rangle \\
			 \end{array}
			 & \left(
			 \begin{array}{cc}
			 	0 & 0 \\
			 	0 & 1
			 \end{array}
			 \right)_{ph}
			\end{array}
			\otimes\begin{array}{c@{\hspace{-5pt}}l}
			 \begin{array}{c}
			 	|0\rangle \\
			 	|1\rangle \\
			 \end{array}
			 & \left(
			 \begin{array}{cc}
			 	1 & 0 \\
			 	0 & 1
			 \end{array}
			 \right)_{at}
			\end{array}
			+\hbar\omega\begin{array}{c@{\hspace{-5pt}}l}
			 \begin{array}{c}
			 	|0\rangle \\
			 	|1\rangle \\
			 \end{array}
			 & \left(
			 \begin{array}{cc}
			 	1 & 0 \\
			 	0 & 1
			 \end{array}
			 \right)_{ph}
			\end{array}
			\otimes\begin{array}{c@{\hspace{-5pt}}l}
			 \begin{array}{c}
			 	|0\rangle \\
			 	|1\rangle \\
			 \end{array}
			 & \left(
			 \begin{array}{cc}
			 	0 & 0 \\
			 	0 & 1
			 \end{array}
			 \right)_{at}
			\end{array}\\
			&+g\left\{\left(\begin{array}{c@{\hspace{-5pt}}l}
			 \begin{array}{c}
			 	|0\rangle \\
			 	|1\rangle \\
			 \end{array}
			 & \left(
			 \begin{array}{cc}
			 	0 & 0 \\
			 	1 & 0
			 \end{array}
			 \right)_{ph}
			\end{array}
			\otimes\begin{array}{c@{\hspace{-5pt}}l}
			 \begin{array}{c}
			 	|0\rangle \\
			 	|1\rangle \\
			 \end{array}
			 & \left(
			 \begin{array}{cc}
			 	1 & 0 \\
			 	0 & 1
			 \end{array}
			 \right)_{at}
			\end{array}\right)\left(\begin{array}{c@{\hspace{-5pt}}l}
			 \begin{array}{c}
			 	|0\rangle \\
			 	|1\rangle \\
			 \end{array}
			 & \left(
			 \begin{array}{cc}
			 	1 & 0 \\
			 	0 & 1
			 \end{array}
			 \right)_{ph}
			\end{array}
			\otimes\begin{array}{c@{\hspace{-5pt}}l}
			 \begin{array}{c}
			 	|0\rangle \\
			 	|1\rangle \\
			 \end{array}
			 & \left(
			 \begin{array}{cc}
			 	0 & 1 \\
			 	0 & 0
			 \end{array}
			 \right)_{at}
			\end{array}\right)\right.\\
			&\left.+\left(\begin{array}{c@{\hspace{-5pt}}l}
			 \begin{array}{c}
			 	|0\rangle \\
			 	|1\rangle \\
			 \end{array}
			 & \left(
			 \begin{array}{cc}
			 	0 & 1 \\
			 	0 & 0
			 \end{array}
			 \right)_{ph}
			\end{array}
			\otimes\begin{array}{c@{\hspace{-5pt}}l}
			 \begin{array}{c}
			 	|0\rangle \\
			 	|1\rangle \\
			 \end{array}
			 & \left(
			 \begin{array}{cc}
			 	1 & 0 \\
			 	0 & 1
			 \end{array}
			 \right)_{at}
			\end{array}\right)\left(\begin{array}{c@{\hspace{-5pt}}l}
			 \begin{array}{c}
			 	|0\rangle \\
			 	|1\rangle \\
			 \end{array}
			 & \left(
			 \begin{array}{cc}
			 	1 & 0 \\
			 	0 & 1
			 \end{array}
			 \right)_{ph}
			\end{array}
			\otimes\begin{array}{c@{\hspace{-5pt}}l}
			 \begin{array}{c}
			 	|0\rangle \\
			 	|1\rangle \\
			 \end{array}
			 & \left(
			 \begin{array}{cc}
			 	0 & 0 \\
			 	1 & 0
			 \end{array}
			 \right)_{at}
			\end{array}\right)\right\}\\
			&=\begin{array}{c@{\hspace{-5pt}}l}
			 \begin{array}{c}
			 	|0\rangle|0\rangle \\
			 	|0\rangle|1\rangle \\
			 	|1\rangle|0\rangle \\
			 	|1\rangle|1\rangle \\
			 \end{array}
			 & \left(
			 \begin{array}{cccc}
			 	0 & 0 & 0 & 0 \\
			 	0 & \hbar\omega & g & 0\\
			 	0 & g & \hbar\omega & 0\\
			 	0 & 0 & 0 & 2\hbar\omega \\
			 \end{array}
			 \right)_{ph\otimes at}
			\end{array}
		\end{aligned}
	\end{equation}
	where $I_a,\ I_{\sigma}$ are unit operators, $\omega=\omega_c=\omega_a$. Via tensor product (shown in Fig. \ref{appxfig:TensorGenerator}(a)) we create a 4 by 4 matrix. Now we assume that initial state is $|0\rangle|1\rangle$ and dissipation of photon is allowed, thus we only have three states $|0\rangle|0\rangle,\ |1\rangle|0\rangle,\ |0\rangle|1\rangle$ in the system. $|1\rangle|1\rangle$ does not exist according to the initial state. It is useless. Hamiltonian is rewritten as
	\begin{equation}
		\label{appxeq:GeneratorAlgoritm1}
		H_{TC}^{RWA}=\begin{array}{c@{\hspace{-5pt}}l}
			 \begin{array}{c}
			 	|0\rangle|0\rangle \\
			 	|0\rangle|1\rangle \\
			 	|1\rangle|0\rangle \\
			 \end{array}
			 & \left(
			 \begin{array}{ccc}
			 	0 & 0 & 0 \\
			 	0 & \hbar\omega & g \\
			 	0 & g & \hbar\omega \\
			 \end{array}
			 \right)_{ph\otimes at}
			\end{array}
	\end{equation}
	Now the new Hamiltonian is 3 by 3. If closed system is considered, then dissipation is forbidden. $|0\rangle|0\rangle$ is also useless. Hamiltonian is as follows
	\begin{equation}
		\label{appxeq:GeneratorAlgoritm2}
		H_{TC}^{RWA}=\begin{array}{c@{\hspace{-5pt}}l}
			 \begin{array}{c}
			 	|0\rangle|1\rangle \\
			 	|1\rangle|0\rangle \\
			 \end{array}
			 & \left(
			 \begin{array}{cc}
			 	\hbar\omega & g \\
			 	g & \hbar\omega \\
			 \end{array}
			 \right)_{ph\otimes at}
			\end{array}
	\end{equation}
	Now the new Hamiltonian is 2 by 2.
	
	\begin{figure}
		\centering
		\includegraphics[width=0.4\textwidth]{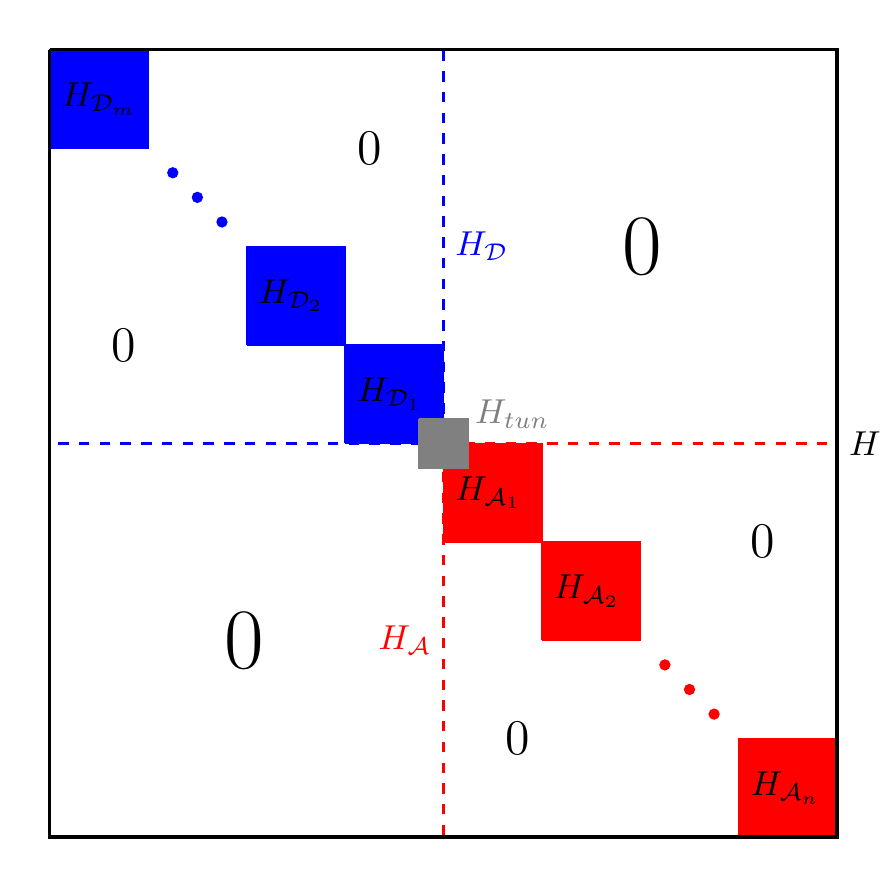}
		\caption{(online color) {\it Hamiltonian of the association-dissociation of neutral hydrogen molecule.} $H=H_{\mathcal{D}}+H_{\mathcal{A}}+H_{tun}=\sum_i^mH_{\mathcal{D}_i}+\sum_j^nH_{\mathcal{A}_j}+H_{tun}$.}
		\label{appxfig:Hamil}
	\end{figure}
	
	Tensor product is actually a common but ineffective method of producing a full Hamiltonian. We typically do not require the complete Hamiltonian due to initial condition restrictions. The relevant portion frequently only takes up a very small portion of the Hilbert space, particularly for complex multi-particle systems. In order to throw away extraneous unneeded states while maintaining beneficial states, we add the generator method depicted in Fig. \ref{appxfig:TensorGenerator}(b). The core is establishing Hamiltonian with states generated by generator algorithm according to initial state and possible interactions and dissipations among them.
	
	The fact that the generator approach does not require an operator is the main distinction between it and the tensor product algorithm. As a result, we are free to assign any number to each state when breaking up the Hamiltonian. For instance, the association--dissociation Hamiltonian of the neutral hydrogen molecule, described in Eq. \eqref{eq:Hamil}, is divided similarly in this article as seen in Fig. \ref{appxfig:Hamil}.
	
	\section{Abbreviations and notations}
	\label{appx:AbbreviationsNotations}
	
	See Tab. \ref{appx:Table}.
		
	\begin{longtable}{ll}
		\caption{List of abbreviations and notations used in this paper.}\\
		\label{appx:Table}\\
		\hline
		\multicolumn{1}{l}{\textbf{Abbreviations/Notations}} & \multicolumn{1}{l}{\textbf{Descriptions}}\\
		\hline
		\endfirsthead
		
		\multicolumn{2}{c}{{\bfseries -- continued from previous page}}\\
		\hline
		\multicolumn{1}{l}{\textbf{Abbreviations/Notations}} & \multicolumn{1}{l}{\textbf{Descriptions}}\\
		\hline
		\endhead
		
		\hline
		\multicolumn{1}{l}{{Continued on next page}}
		\endfoot
		
		\hline
		\endlastfoot
		
		QED & Quantum electrodynamics\\
		SC & Strong coupling\\
		USC & Ultrastrong-coupling\\
		DSC & Deep strong coupling\\
		QRM & Quantum Rabi model\\
		JCM & Jaynes--Cummings model\\
		TCM & Tavis--Cummings model\\
		JCHM & Jaynes--Cummings--Hubbard model\\
		TCHM & Tavis--Cummings--Hubbard model\\
		QME & Quantum master equation\\
		RWA & Rotating wave approximation\\
		AO & Atomic orbital\\
		MO & Molecular orbital\\
		$ph$ & Photon\\
		$e$ & Electron\\
		$at$ & Atom\\
		$or$ & Orbital\\
		$n$ & Nucleus\\
		$s$ & Spin\\
		$\uparrow$ & Spin up\\
		$\downarrow$ & Spin down\\
		$cb$ & Covalent bond\\
		$\Phi_0$ & Bonding orbital or molecular ground orbital\\
		$\Phi_1$ & Antibonding orbital or molecular excited orbital\\
		$\eta$ & Maximum ratio of coupling strength to frequency\\
		$\omega_c$ & Cavity frequency or photonic mode\\
		$\omega_n$ & Transition frequency, including in molecular (or as $\omega$) and in atom (or as $\Omega$)\\
		$\omega$ & Transition frequency for electron in molecule (e.g. $\omega^{\uparrow},\ \omega^{\downarrow}$)\\
		$\omega^{\uparrow}$ & Transition frequency for electron with $\uparrow$ in molecule\\
		$\omega^{\downarrow}$ & Transition frequency for electron with $\downarrow$ in molecule\\
		$\Omega$ or $\omega_a$ & Transition frequency for electron in atom (e.g. $\Omega^{\uparrow},\ \Omega^{\downarrow}$)\\
		$\Omega^{\uparrow}$ & Transition frequency for electron with $\uparrow$ in atom\\
		$\Omega^{\downarrow}$ & Transition frequency for electron with $\downarrow$ in atom\\
		$\Omega^s$ & Electron spin transition frequency in atom\\
		$\Omega^c$ & Phonon mode\\
		$\mathcal{C}$ & Space of quantum states for entire system\\
		$\mathcal{A}$ & Subspace of quantum states for associative system (or molecular system)\\
		$\mathcal{D}$ & Subspace of quantum states for dissociative system (or atomic system)\\
		$\rho$ & Density Matrix\\
		$\mathcal{L}\left(\rho\right)$ & Lindblad superoperator\\
		$\mathcal{K}$ & Graph of the potential photon dissipations between the states that are permitted\\
		$\mathcal{K}'$ & Graph of the potential photon influxes between the states that are permitted\\
		$L_k\left(\rho\right)$ & Standard dissipation superoperator\\
		$L_{k'}\left(\rho\right)$ & Standard influx superoperator\\
		$\gamma_{k}$ & Total spontaneous emission rate for photon\\
		$\gamma_{k'}$ & Total spontaneous influx rate for photon\\
		$\mu$ & Ratio of influx rate to emission rate (e.g. $\mu_{\omega},\ \mu_{\Omega},\ \mu_{\Omega^s}$)\\
		$A_k$ & Lindblad or jump operator of system and its hermitian conjugate operator --- $A_k^{\dag}$\\
		$H$ & Hamiltonian\\
		$\hbar$ & Reduced Planck constant or Dirac constant\\
		$a$ & Photon annihilation operator (e.g. $a_{\omega},\ a_{\Omega},\ a_{\Omega^s}$) and its hermitian conjugate operator --- $a^{\dag}$\\
		$\sigma$ & \tabincell{l}{Interaction operator of atom with the electromagnetic field of the cavity (e.g. $\sigma_{\omega},\ \sigma_{\Omega},\ \sigma_{\Omega^s}$,\\$\sigma_n$) and its hermitian conjugate operator --- $\sigma^{\dag}$}\\
		$g$ or $g_n$ & Coupling strength of photon and the electron (e.g. $g_{\omega},\ g_{\Omega}$)\\
		$\zeta$ & Nucleus tunnelling strength or atom leap strength (e.g. $\zeta_0,\ \zeta_1,\ \zeta_2$)\\
		$\mathcal{G}\left(T\right)_f$ & Thermally stationary state\\
		$K$ & Boltzmann constant\\
		$T$ & Temperature for photonic mode (e.g. $T_{\omega},\ T_{\Omega},\ T_{\Omega^s}$)\\
		$c$ & Normalization factors (e.g. $c_0,\ c_1,\ c_2,\ c_3$)\\
		$V$ & Effective volume of the cavity\\
		$d$ & Dipole moment of the transition between the ground and the perturbed statesy\\
		$E\left(x\right)$ & Spatial arrangement of the atom in the cavity\\
		$l$ & Length of the cavity\\
		$\lambda$ & Photon wavelength\\
		$N$ & Number of atoms\\
		$M$ & Number of cavities\\
		$I_a$ & Unit operator\\
		$I_{\sigma}$ & Unit operator\\
	\end{longtable}
	
\end{document}